\documentclass[11pt,preprint,aps,prd]{revtex4-2}
\usepackage{silence}
\usepackage[utf8]{inputenc}
\usepackage{textcomp, gensymb}
\usepackage{csvsimple}
\usepackage{amsmath}
\usepackage{amssymb}
\usepackage{hyperref}
\usepackage{graphicx}
\graphicspath{{./images/}}
\usepackage[caption=false]{subfig}
\usepackage{float}
\usepackage{booktabs}
\usepackage{bigstrut}
\usepackage{gensymb}
\usepackage{geometry}
\usepackage{physics}
\usepackage{leftindex}
\usepackage{bm}
\usepackage[dvipsnames]{xcolor}

\usepackage{subfig}
\usepackage{comment}
\usepackage[most]{tcolorbox}

\def\f{\frac}

\def\mc{\mathcal}

\newcommand{\lreg}{\ell_{\text{reg}}}
\newcommand{\rhoconst}{\gamma}
\newcommand{\rholens}{\rho_{\text{lens}}}
\newcommand{\nn}{\nonumber}

\begin{document}
\title{Self-force and the Schwarzschild star}

\author{Abhinove N.~Seenivasan}
\affiliation{
Consortium for Fundamental Physics,
School of Mathematical and Physical Sciences,
University of Sheffield,
Hicks Building,
Hounsfield Road,
Sheffield S3 7RH, 
United Kingdom.}

\author{Sam R.~Dolan}
\affiliation{
Consortium for Fundamental Physics,
School of Mathematical and Physical Sciences,
University of Sheffield,
Hicks Building,
Hounsfield Road,
Sheffield S3 7RH, 
United Kingdom.}

\begin{abstract}
We consider the self-force acting on a pointlike (electromagnetic or conformal-scalar) charge held fixed on a spacetime with a spherically-symmetric mass distribution of constant density (the Schwarzschild star).  The Schwarzschild interior is shown to be conformal to a three-sphere geometry; we use this conformal symmetry to obtain closed-form expressions for mode solutions. We calculate the self-force with two complementary regularization methods, direct and difference regularization, showing agreement. For the first time, we show that difference regularization can be applied in the non-vacuum interior region, due to the vanishing of certain regularized mode sums.  The new results for the self-force come in three forms: series expansions for the self-force near the centre of the star and in the far field; a new approximation that describes the divergence in the self-force near the star's boundary; and numerical data presented in a selection of plots. We conclude with a discussion of the logarithmic divergence in the self-force in the approach to the star's surface, and the effect of boundaries.
\end{abstract}

\maketitle
\newpage


\section{Introduction}

An electric charge interacts with electromagnetic fields in its vicinity. Since an electric charge is also the source for an electromagnetic field, one cannot avoid the question of self-interactions, even in  classical field theory. For example, it is well-known that an accelerated charge generates radiation. This is accompanied by a \emph{radiation reaction} force on the particle which ensures that energy is properly conserved. The earliest such inquiries led to the Abraham-Lorentz-Dirac force law \cite{doi:10.1098/rspa.1938.0124}, and this has been an active field of enquiry ever since.

The presence of spacetime curvature arising in General Relativity (GR) adds to the subtlety of the self-interaction problem. Even in the simple case of a static electric charge, the field lines are not completely isotropic in the vicinity of a pointlike particle, due to the interplay between the electromagnetic field and the background curvature. In other words, one cannot avoid the notion of a self force (SF) \cite{DEWITT1960220} even in elementary scenarios. The theoretical arguments that lead to the requirement of a self-force hold equally well for gravitational fields generated by masses, as well as for scalar fields generated by scalar charges. The latter, being the simplest case of all, typically serves as an instructive model for understanding general aspects of the theory. The calculation of the SF for scalar, electric and massive particles has been the focus of the community for over three decades and reviews of the subject can be found in \cite{Poisson:2011nh} and \cite{Barack:2018yvs}. 

The gravitational self-force is of particular importance because it is used to model the orbital evolution of binary systems with a small mass ratio \cite{Barack_2010, Diener_2012, Warburton_2012}, that are radiating energy in the form of gravitational waves (GW). The scalar and EM self-forces are not of such immediate experimental relevance, it would appear, but nevertheless they are interesting in their own right, and they serve to clarify aspects that are more opaque in the former case (for example, unlike the gravitational self-force, the scalar and EM self-forces are not gauge-dependent). We focus on the latter cases here.

The electromagnetic self-force of charged particles has been explored since the work of Copson \cite{copson1928electrostatics} in the 1920s. In the 1980s, Smith and Will \cite{Smith:1980tv} showed that a pointlike particle of electromagnetic charge $q$ held fixed at $r = r_0$ outside a Schwarzschild black hole (BH) of mass $M$, experiences a self-force that is \emph{repulsive}, and equal in magnitude to $q^2 M / r_0^3$. Here we have adopted units such that $G = c = 4 \pi \epsilon_0 =1$. 

For scalar fields, a key result, due to Wiseman \cite{Wiseman:2000rm}, is that the SF on a scalar charge held stationary outside a Schwarzschild black hole (BH) is precisely zero. Conversely, outside a Newtonian star (i.e.~a dilute star with $R \gg M$), Poisson and Pfenning \cite{Pfenning:2000zf} showed that a scalar charge held static at radius $r$ experiences a self-force of $F_r = 2 \xi q^2 M / r^3$, where $\xi$ is the coupling to the Ricci scalar (with $\xi = 1/6$ for conformal coupling and $\xi = 0$ for minimal coupling). Taken together, these results indicate that the self-force depends on the global structure of the spacetime, and not just on its local geometry.

Calculations of the self-force for pointlike particles, such as those above, typically involve \emph{regularization}, whereby a symmetric (i.e.~isotropic around the particle) but formally singular part of the field is removed to leave a regular remainder. Regularization is necessary (but straightforward) even in flat spacetime, when dealing with pointlike particles. In the simple example of a pointlike charge $q$ held at $z=d$ above a grounded conducting plate at $z=0$, the electric potential is (by the method of image charges) given by $V = V_S + V_R$ where
\begin{equation}
  V_S \equiv  \frac{q}{\sqrt{x^2 + y^2 + (z-d)^2}} , \quad 
  V_R \equiv -\frac{q}{\sqrt{x^2 + y^2 + (z+d)^2}} . 
\end{equation}
Before one can compute the force on the pointlike charge at $z=d$ it is necessary to regularize by removing the term $V_S$. The force on the particle is then given by $\mathbf{F} = -q \boldsymbol{\nabla} V_R = - \frac{q^2}{4 d^2} \hat{\mathbf{z}}$. In this example it is notable that this (self) force is attractive, proportional to $q^2$, and clearly dependent on boundary conditions. Moreover, this force diverges as $d \rightarrow 0$. In this work, we shall show that the self-force in curved spacetime also diverges, albeit in weaker fashion, at the boundary of a Schwarzschild star.

In vacuum regions of spacetimes of spherical symmetry, one can circumvent the need for removing the S field directly by calculating \emph{self-force differences}, following the prescription of Drivas \& Gralla \cite{Drivas:2010wz} (see also Isoyama \& Poisson \cite{Isoyama_2012}). As known from Birkhoff's theorem, the spacetime in the vacuum region outside an (isolated) spherically-symmetric body is exactly the Schwarzschild solution with the same mass. The symmetric/singular `S' part of the particle's field depends \emph{only} on the local geometry, and so it is the same in the two spacetimes for a particle held at the same radius; hence taking the difference of the fields eliminates the singular part, effectively regularizing the field. In this work, we compute self-forces both with direct methods (i.e.~with a mode-sum regularization step) and with this method of differences, arriving at self-consistent results. Intriguingly, we find that the difference method can also be applied inside the star, a non-vacuum region where Birkhoff's theorem does not apply.

In principle, the SF can be used to distinguish the presence of a compact body from that of a BH of the same mass. An interesting subtlety is that the force on a charge that is \emph{far} from the BH or compact body has a \emph{universal character} at leading order. More precisely, the SF for \emph{minimally-coupled} (but not conformally-coupled) scalar fields, and electromagnetic fields, is agnostic to the spacetime topology and mass distribution at leading order in the $M/r_0$ expansion \cite{Shankar_2007, Drivas:2010wz, Isoyama_2012}. At higher orders, one finds structure-dependent factors (for example, coefficients sensitive to $M/R$, where $R$ is the body's radius) which distinguish the respective forces outside compact objects and BHs of the same mass. In short, despite universality at leading order, the global spacetime structure \emph{can} be probed using the SF. 

In this paper, we obtain new results for the self-force of a charge held at fixed position in the vicinity of a Schwarzschild star of radius $R$ (i.e.~a matter distribution of constant density) in three forms. First, as series expansions, in the far-field ($r \gg R$), and near the centre of the star; second, as an approximation near the surface of the star that encapsulates a logarithmic divergence; and, third, as numerical data and plots.

This paper is organised as follows. In Sec.~\ref{sec:scalar}, we review methods for calculating self-force for (static, conformally-coupled) scalar fields on spherically-symmetric spacetimes via regularized mode expansions. In Sec.~\ref{sec:star}, we show that the Schwarzschild star interior geometry is conformal to a three-sphere geometry, which leads to simple closed-form expressions for mode solutions. In Sec.~\ref{sec:EMfield}, we review the formulation of the \emph{electromagnetic} self force, and in Sec.~\ref{sec:identities} we describe a method for obtaining exact results from certain regularized mode sums. The results section (Sec.~\ref{sec:results}) includes the verification of the two regularization methods (Sec.~\ref{sec:regforces}); series expansions for the SF in the far-field and near the star's centre (Sec.~\ref{sec:series}); a closer look at the divergence of the SF in the approach to the star's surface (Sec.~\ref{sec:boundary}); and finally plots of the self-force across the full domain in $r$ (Sec.~\ref{sec:full-domain}). We conclude in Sec.~\ref{sec:conclusions} with a discussion of the key results. A Mathematica notebook with codes to generate plots and compute the SF can be found in \cite{Mathematica-Notebook}.

\section{Formulation\label{sec:formulation}}

 \subsection{Scalar fields: Self force, mode expansions and regularization\label{sec:scalar}} 

 \subsubsection{Spherically-symmetric spacetimes}
We will work with static, spherically symmetric spacetimes whose line elements can be written in the form
\begin{equation}\label{SSSS}
    ds^2 = -A(r) dt^2 + \frac{1}{B(r)} dr^2 + r^2 d\Omega^2~, 
\end{equation}
where $d\Omega^2$ is the metric on a unit 2-sphere. We use a metric signature with the ``mostly positive" sign convention $(-+++)$, and geometric units such that $G = c = 4 \pi \epsilon_0 = 1$. We will work with the Schwarzschild exterior and interior spacetimes, with the latter modelling the interior of a (neutral, transparent) `star' of constant density.
 
  \subsubsection{Scalar field equations} 
A scalar charge $q$ interacting with a scalar field $\varphi$ that is non-minimally coupled to gravity with coupling $\xi$ is modelled with the action
\begin{align}
    \mc{S} &= \mc{S}_{\text{F}} + \mc{S}_{\text{P}} + \mc{S}_{\text{Int}}~, \label{scaction} \\
     &= -\frac{1}{8\pi}\int\sqrt{-g}~d^4x\left\{{g^{\mu \nu}}\nabla_\mu\varphi\nabla_\nu\varphi + \xi \mc{R} \varphi^2\right\} -m_0\int d\tau +q\int_\gamma \sqrt{-g}~d^4x~\varphi(x)~\delta^4(x,x_0)~d\tau~,
\end{align}
where $\mc{S}_{\text{Int}}$ is defined with reference to a particle worldline $\gamma : x_0^\mu(\tau)$. Here, $\mc{R}$ is the Ricci scalar and $m_0$ is the bare mass of the particle. The three terms in (\ref{scaction}) are the field action, the particle action and the interaction term, respectively. 
The dynamical equation for the scalar field is 
\begin{equation}\label{KG}
    \left(\Box - \xi \mc{R} \right)\varphi = -4 \pi \rho, \quad \quad 
    \rho \equiv q\int_\gamma \delta^4(x,x_0(\tau))~d\tau~, 
\end{equation}
where $\delta^4(x,x_0(\tau))$ is the Dirac distribution. The source term can be evaluated by performing the integral over $t$ to obtain 
\begin{equation}
    \rho = \f{q}{u^t}\f{\delta^3(x^i- x^i_0)}{\sqrt{-g}}~,
\end{equation}
where $u^t = dt / d\tau$ is the Lorentz factor for the static particle.

 \subsubsection{Mode expansions}
One can expand the field $\varphi$ in spherical harmonics by exploiting the spherical symmetry in the problem, with the following decomposition: 
\begin{equation}\label{fielddecomp}
    \varphi = \sum_{\ell,m}\psi_{\ell m}(r)Y_{\ell m}(\theta, \phi) . 
\end{equation}
The Dirac distribution can also be expanded in spherical harmonics, as 
\begin{equation}\label{DiracYLM}
    \f{\delta^3(x^i- x^i_0)}{\sqrt{-g}} = \sqrt{\f{B}{A}}\f{\delta(r-r_0)}{r^2}\sum_{\ell,m}Y_{\ell m}(\theta, \phi)Y^{*}_{\ell m}(\theta_0, \phi_0)~.
\end{equation}
Without loss of generality, one can assume that the static charge is on the $z$ axis at $\theta_0 = 0$ and thus use $Y_{\ell m}(0,\phi) = \sqrt{\f{2\ell+1}{4\pi}}\delta_{m,0}$ and hence $\psi_{\ell m}(r) = \psi_{\ell}(r) \delta_{m,0}$.

Through this separation of variables and mode sum, the PDE (\ref{KG}) is reduced to (a set of) ODEs in the radial variable $r$. After introducing a tortoise coordinate $r_*$ defined by $dr/dr_* = \sqrt{AB}$, the radial equation becomes
\begin{equation}\label{genscalartortoise}
    \dv[2]{u_{\ell}}{r_*} - V(r)u_{\ell} =  - 4\pi\f{q}{r_0} \sqrt{A_0}\sqrt{\f{2\ell+1}{4\pi}} \delta\left(r^*-r^*_0\right)~,
\end{equation}
with $u_{\ell}(r) \equiv r \psi_{\ell}(r)$, $A_0 \equiv A(r_0)$, $B_0 \equiv B(r_0)$ and the effective potential
\begin{equation}
    V(r) = \f{A'B+AB'}{2 r} + A\left(\f{\ell(\ell+1)}{r^2} + \xi R\right)~,
\end{equation}
where the prime ${}^\prime$ indicates a derivative with respect to $r$. 

\subsubsection{Radial solutions and self-force}
To proceed further with standard methods, one needs homogeneous solutions to the ODE that satisfy the physical boundary conditions. For now, let us assume that we have two such homogeneous linearly independent solutions. The solution that satisfies the physical boundary conditions at the horizon (in the case of a BH) or at $r \leq r_0$ is termed the ``in" solution while the solution that satisfies the boundary conditions at infinity is termed the ``up" solution, and we denote these by $\psi_{\ell}^{\text{in}}(r)$ for $r \leq r_0$ and $\psi_{\ell}^{\text{up}}(r)$ for $r \geq r_0$, with $u^{\text{in/up}}_{\ell}(r) \equiv r \psi^{\text{in/up}}_{\ell 0}(r)$. 

The radial equation \eqref{genscalartortoise} implies that $u_\ell(r)$ is continous at $r_0$ and that there is a discontinuity in its first derivative. Hence we adopt the standard ansatz 
\begin{equation}\label{genscalaransatz}
    u_{\ell}(r) =  \mc{N}_{\ell}\begin{cases}
        u_{\ell}^{\text{in}}(r)u_{\ell}^{\text{up}}(r_0)~, & r \leq r_0~, \\
        u_{\ell}^{\text{in}}(r_0)u_{\ell}^{\text{up}}(r)~, & r \geq r_0~,
     \end{cases}
\end{equation}
where $\mc{N}_{\ell}$ is a constant to be determined. Integrating over the source in Eq.~\eqref{genscalartortoise} yields the junction condition
\begin{equation}\label{junctioncond}
\lim_{\epsilon \rightarrow 0^+} \left( 
  \left. \dv{u_{\ell}}{r_*} \right|_{r_0+\epsilon} 
- \left. \dv{u_{\ell}}{r_*} \right|_{r_0-\epsilon} \right)
  =   - 4\pi\f{q}{r_0} \sqrt{A_0}\sqrt{\f{2\ell+1}{4\pi}}~,
\end{equation}
from which it follows that
\begin{align}
    \mc{N}_{\ell} &= - \f{4\pi}{\sqrt{B_0}}\f{q}{r_0^3}\sqrt{\f{2\ell+1}{4\pi}}\f{1}{\mc{W}_{\ell}}~,
\end{align}
where $\mc{W}_{\ell}$ is the Wronskian for the $\psi_{\ell}$ solutions, given by $\mc{W}_{\ell} = \psi_{\ell}^{\text{in}}\left(\psi_{\ell}^{\text{up}}\right)'-\psi_{\ell}^{\text{up}}\left(\psi_{\ell}^{\text{in}}\right)'$.

\subsubsection{Self-force: $\ell$-modes}
The `bare' $\ell$-modes of the radial SF \emph{before regularization} may be obtained through a limit
\begin{equation}
 F_r^{\ell \pm} = q Y_{\ell 0}(0) \lim_{\epsilon \rightarrow 0^\pm} \partial_r \psi_\ell(r_0 + \epsilon) .
\end{equation}
However, as noted in Eq.~\eqref{junctioncond}, the radial derivative of the $\ell$ modes is not continuous at $r = r_0$, and so $F_r^{\ell+} \neq F_r^{\ell-}$. For our purposes, it is convenient to work with the average, given by
\begin{equation} \label{eq:Fl-bare}
    F_r^{\ell,\text{bare}} \equiv \frac{1}{2} \left(F_r^{\ell+} + F_r^{\ell-} \right) = -\left(\f{q}{r_0}\right)^2\f{\left(2\ell+1\right)}{2\mc{W}_{\ell}\sqrt{B_0}} \left\{\dv{\psi_{\ell}^{\text{in}}}{r}\psi_{\ell}^{\text{up}} + \psi_{\ell}^{\text{in}}\dv{\psi_{\ell}^{\text{up}}}{r}\right\} ~,
\end{equation}
where all quantities are now evaluated at the source point $r_0$. 

Regularization of the self-force can be achieved through two  approaches, namely, \emph{direct} regularization and \emph{difference} regularization, which we now describe.

\subsubsection{Direct regularization}\label{directmethod}

The canonical method of regularising the SF, as formalised by Detweiler and Whiting \cite{Detweiler_2003}, involves decomposing the Green's functions  
for the relevant field equations into \textit{singular} (S) and \textit{regular} (R) parts in a unique way. These S and R Green's functions are then used to define S and R fields, such that we have a split $\Phi_{\text{ret}} = \Phi_S + \Phi_R$. The S field $\Phi_S$ is symmetric around the charge, and has the same singularity structure as  $\Phi_{\text{ret}}$ at the particle's position. The S field, which has an entirely local construction, does not contribute to the self force. The R field is the difference between the retarded and S fields ($\Phi_R = \Phi_{\text{ret}} - \Phi_S$); it is a solution to the source-free equations of motion (i.e.~$\Box \Phi_R = 0$) and it is wholly responsible for the SF:
\begin{equation}
F_\mu = q \nabla_\mu \Phi_R .
\end{equation}
The Detweiler-Whiting regularisation method is general; it is not restricted to the case of static or spherically symmetric spacetimes \cite{Harte_2006, Harte_2009, Harte_2012, PhysRevD.81.024023}. 
However, in the static scenario, only the radial component of the force is non-zero, due to the symmetry of the setup.

Given a mode-sum decomposition for the SF, a practical approach to regularisation is to subtract from each `bare' $\ell$-mode a corresponding contribution that arises from a multipole decomposition of the gradient of the S field in the vicinity of the charge. The (regularized) self-force is 
\begin{equation}
 F_\mu = \sum_{\ell} F_\mu^{\ell,\text{reg}}
\end{equation}
where
\begin{equation} \label{eq:Fr-regularized}
    F_r^{\ell,\text{reg}} = F_r^{\ell\pm} - q^2\left\{ \widetilde{A}^\pm\left(\ell +\f{1}{2}\right) + \widetilde{B} + \f{\widetilde{C}}{\left(\ell + \f{1}{2}\right)} + \f{\widetilde{D}}{\left(\ell - \f{1}{2}\right)\left(\ell + \f{3}{2}\right)} + \ldots \right\}~,
\end{equation}
and $\widetilde{A}^\pm,\widetilde{B},\widetilde{C},\widetilde{D}, \ldots$ are known as \emph{regularisation parameters} \cite{PhysRevD.61.061502, Poisson:2011nh}. The values that the parameters take depends on the the motion of the pointlike charge, the background spacetime and the nature of the field under consideration. 

For static \emph{minimally coupled} scalar fields in static, spherically symmetric spacetimes, the regularisation parameters were calculated by Casals, Poisson and Vega in Ref.~\cite{PhysRevD.86.064033} (see also \cite{Burko_1999, PhysRevD.61.061502} for earlier work) to be
\begin{align}\label{scalarregparam}
    \widetilde{A}^\pm = \mp \f{1}{r^2}B^{-1/2}, &\quad \widetilde{B} = - \f{1}{2r^2}\left(1+r\Psi'\right), \quad \widetilde{C} = 0, \\
    \widetilde{D} = -\f{1}{16r^2}\left[\left(1+r\Psi'\right) - \right.&\left.\left(1+r\Psi' + 3r^2\Psi'^2 + r^3\Psi'^3 - 6r^2\Psi'' - 2r^3\Psi'''\right)B +\right. \\
    &\left.\left(1+4r\Psi' +3r^2\Psi''\right)rB'+\left(1+r\Psi'\right)r^2B''\right]~,
\end{align}
where $A(r) = e^{2\Psi}$ and $B = B(r)$ are the metric functions in Eq.~\eqref{SSSS}, and $B'(r) = \frac{dB}{dr}$, etc. Since $\widetilde{A}^+ = -\widetilde{A}^-$, it is convenient to regularize the \emph{averaged} bare modes, as follows:
\begin{equation}
    F_r^{\ell,\text{reg}} = F_r^{\ell, \text{bare}} - q^2\left\{ \widetilde{B} + \f{\widetilde{C}}{\left(\ell + \f{1}{2}\right)} + \f{\widetilde{D}}{\left(\ell - \f{1}{2}\right)\left(\ell + \f{3}{2}\right)} + \ldots \right\}~.
\end{equation}
For static \emph{conformally coupled} fields (considered here), the $\widetilde{A}^\pm$, $\widetilde{B}$ and $\widetilde{C}$ terms are identical to the above, but the $\widetilde{D}$ term differs in non-vacuum regions (i.e.~in the interior of the star).

\subsubsection{Difference regularization}\label{diffmethod}

The second approach we use is the \textit{difference method} pioneered by Drivas and Gralla \cite{Drivas:2010wz}, and developed and applied by Isoyama and Poisson \cite{Isoyama_2012}. In its standard form, it relies on Birkhoff's theorem, and so it has only been used to evaluate the self-force at a point in a vacuum region under the assumption of spherical symmetry. Since the S-field has a local construction, the S-field in the vacuum exterior of the star is equal to the S-field in the BH Schwarzschild spacetime at the same radius $r_0$. Thus, the difference in the retarded fields must equal the difference in the R fields, and hence one can compute the \textit{SF difference} with solutions in the two spacetimes. Additionally, in the case of the static scalar fields the BH SF \emph{vanishes} \cite{Wiseman:2000rm}, and so the difference is equal to the full SF. 

Applying the discussion in the previous section to the case of the Schwarzschild BH, for which $A(r) = 1 - 2M /r = B(r)$, one has the exterior homogeneous solutions in terms of the Legendre functions, i.e.~$\psi_\ell^{\text{in}}(r) = P_{\ell}(z)$ and $\psi_\ell^{\text{up}}(r) = Q_{\ell}(z)$, where we have introduced the harmonic coordinate 
\begin{equation}
z = r/M - 1.  \label{eq:zdef}
\end{equation}
The Wronskian for the Legendre polynomials is simply $\mc{W}_{\ell} = -(z^2-1)^{-1}$. For the black hole case, the bare modes of the force (see Eq.~\eqref{eq:Fl-bare}) are then
\begin{equation} \label{eq:FBH}
    \left(F_r^l\right)^{BH}=\f{1}{2} \left(\f{q}{M}\right)^2\sqrt{\f{z_0-1}{z_0+1}} \left(2\ell+1\right)\left\{P_{\ell}'(z_0) Q_{\ell}(z_0) + Q_{\ell}'(z_0) P_{\ell}(z_0) \right\}~,
\end{equation}
where here the $'$ denotes differentiation with respect to the argument $z$. 

For a field in the exterior spacetime outside a star, one can modify Eq.~(\ref{genscalaransatz})) to write down a general solution $\psi^{\text{ext}}_\ell(z)$ that is compatible with the (outer) boundary condition, viz.,
\begin{equation}\label{stellaransatz}
    \psi^{\text{ext}}_\ell(z) = \begin{cases}
                            A_\ell~P_{\ell}(z) + B_\ell~Q_{\ell}(z), & Z < z < z_0~, \\
                            C_\ell~Q_{\ell}(z), & z > z_0~,
                        \end{cases}
\end{equation}
where $Z = R/M-1$ is the position of the star surface and $z_0 = r_0/M-1$ is the location of the charge in the harmonic coordinate, and $A_\ell$, $B_\ell$, $C_\ell$ are coefficients to be determined. Furthermore, suppose we have an solution in the interior of the star which satisfies the boundary condition at $r=0$, that we can write as $D_\ell \varphi^{\text{int}}_\ell(r)$ for some coefficient $D_\ell$. The interior solution and its normal derivative must be matched at the boundary of the star (at $r=R$) with the external solution in Eq.~\eqref{stellaransatz}. In addition, at $r=r_0$ the $\ell$-mode of the field should be continuous across the source and it should satisfy the junction condition in Eq.~(\ref{junctioncond}). These four conditions are sufficient to determine the four coefficients $A_\ell, B_\ell, C_\ell \text{ and } D_\ell$ uniquely, in principle, for each $\ell$. 

Since matching conditions are applied at two points in the domain, $z = Z$ and $z = z_0$, it is worthwhile to make some re-definitions, following Ref.~\cite{Isoyama_2012}. Let $\alpha_\ell = A_\ell P_{\ell}(z_0), \beta_\ell = B_\ell Q_{\ell}(z_0) \text{ and } \gamma_\ell = C_\ell Q_{\ell}(z_0)$. Further, let $\delta_\ell = D_\ell \varphi^{\text{int}}_\ell(Z)$ and 
\begin{equation} \label{eq:eta}
    \eta_\ell \equiv \left. \frac{d\ln{\varphi^{\text{int}}_{\ell}}}{d\ln{r}} \right|_{r=R}.
\end{equation} 
 Carrying out the matching, the constants are determined to be $\delta_\ell = \alpha_\ell P_{\ell}(Z)/P_{\ell}(z_0) + \beta_\ell Q_{\ell}(Z)/Q_{\ell}(z_0)$ and $\gamma_\ell = \alpha_\ell + \beta_\ell$ where 
\begin{align}
    \alpha_\ell &= \left(z_0-1\right)\left(z_0+1\right)P_{\ell}(z_0)Q_{\ell}(z_0) J_\ell~, \\
    \beta_\ell &= - \left(z_0-1\right)\left(z_0+1\right) S_{\ell} \left[Q_{\ell}(z_0)\right]^2 J_\ell~, \\
    J_\ell &= 4\pi \f{q}{M}\sqrt{\f{2l+1}{4\pi}}\f{1}{\sqrt{\left(z_0-1\right)\left(z_0+1\right)^{3}}}~, \\
    S_{\ell} &= \f{\left(Z+1\right)P_{\ell}'(Z) - \eta_{\ell} P_{\ell}(Z)}{\left(Z+1\right)Q_{\ell}'(Z) - \eta_{\ell} Q_{\ell}(Z)}~.  \label{eq:structure-fac}
\end{align}

Here, $S_{\ell}$ is a \emph{structure factor} that only depends on the star's properties, and not on the position and properties of the charge \cite{Drivas:2010wz,Isoyama_2012}. The solution in the exterior is now determined uniquely. Evaluating the derivative at the particle position yields the bare modes of the self force in the star spacetime, given by
\begin{equation}\label{starSF1}
    \left(F_r^{\ell}\right)^{\text{star}} =  \f{1}{2} \left(\f{q}{M}\right)^2\sqrt{\f{z_0-1}{z_0+1}} \left(2l+1\right)\left\{P_{\ell}'Q_{\ell} + Q_{\ell}'P_{\ell} - 2 S_{\ell} Q_{\ell} Q_{\ell}'\right\}~,
\end{equation}
with all Legendre functions evaluated at $z = z_0$. From comparing the above with Eq.~\eqref{eq:FBH}, one observes that $\ell$-modes of the self-force on a star spacetime are equal to those for the self force on a black hole spacetime, plus an additional `difference' piece. The final step is to regularize the self-force and to sum over $\ell$-modes, as in Sec.~\ref{directmethod}, to obtain 
\begin{equation}\label{starSF2}
F_r^{\text{star}} = F_r^{\text{BH}} + \Delta F^{\text{star}}_r~,
\end{equation} 
where
\begin{align}
    F_r^{\text{BH}} &= \f{1}{2} \left(\f{q}{M}\right)^2\sqrt{\f{z_0-1}{z_0+1}} \sum_{\ell=0}^{\infty} \left( \left(2l+1\right)\left\{P_{\ell}'Q_{\ell} + Q_{\ell}'P_{\ell} \right\} - \widetilde{b}_l \right), \label{eq:FrBHsum} \\
    \Delta F^{\text{star}}_r &= -  \left(\f{q}{M}\right)^2\sqrt{\f{z_0-1}{z_0+1}}\sum_{\ell}\left(2\ell+1\right)S_{\ell} \, Q_{\ell}(z_0) \, Q_{\ell}'(z_0)~. \label{eq:Frdiff}
\end{align}
Here $\Delta F_r^{\text{star}}$ is the \emph{self force difference}, in the form obtained by Isoyama \& Poisson \cite{Isoyama_2012} but with an appropriately-modified structure factor $S_\ell$ for the Schwarzschild star. The rescaled regularization parameter in Eq.~\eqref{eq:FrBHsum} is \begin{equation}
    \widetilde{b}_\ell \equiv 2M^2 \sqrt{\f{z_0+1}{z_0-1}} \widetilde{B}_\ell 
    = \frac{d}{dz_0} \left( \frac{1}{\sqrt{z_0^2 - 1}} \right) , 
    \label{eq:bell-tilde}
\end{equation} 
with $\widetilde{B}_\ell$ as given in Eq.~\eqref{scalarregparam}. 

The regularized mode sum in Eq.~\eqref{eq:FrBHsum} is precisely zero, consistent with a vanishing BH self-force $F_r^{\text{BH}}$ for a static scalar particle derived in Ref.~\cite{Wiseman:2000rm}; we examine this point more closely in Sec.~\ref{sec:identities}. Hence the self-force is given by the difference mode-sum in Eq.~\eqref{eq:Frdiff}, and this sum is exponentially convergent with $\ell$ everywhere in the domain except at the star's surface at $r = R$ (this point is examined in Sec.~\ref{sec:boundary}).

\subsection{The Schwarzschild star and the Maxwell fisheye lens} \label{sec:star}
To make further progress, we require mode solutions on the interior spacetime that are regular at the star's centre (i.e.~at $r=0$). Below we describe the interior spacetime in more detail, and we rewrite it in terms of isotropic coordinates \cite{wyman1946schwarzschild}. This highlights a connection between the interior Schwarzschild metric and the Maxwell fisheye lens (see also Ref.~\cite{xiao2023analogy}), and a link to the three-sphere geometry. This leads us to closed-form expressions for mode functions of (conformally-coupled) fields in the interior region.

 \subsubsection{The Schwarzschild exterior}
The solution of Einstein's field equations under the assumptions of spherical symmetry and vacuum is well-known: it is the Schwarzschild exterior solution, described by the line element of Eq.~\eqref{SSSS} with $A(r) = B(r) = 1 - 2 M / r$, where $M$ is the mass in the spacetime. It is also well-known that, after introducing a new radial coordinate $\rho$, the line element can be written in \emph{isotropic form} as \cite{thorne2000gravitation}
\begin{equation} \label{isotropic-form}
    ds^2 = \alpha^2(\rho) \left[-dt^2 + n^2(\rho) \left(d\rho^2 + \rho^2 d\Omega^2 \right) \right] ,
\end{equation}
with
\begin{align}
 \alpha(\rho) &= \frac{1 - M / (2 \rho)}{1 + M / (2 \rho)},   &
 n(\rho) &= \frac{\left(1 + M / (2\rho)\right)^3}{1 - M/(2\rho)} ,
\end{align}
where
\begin{align}
 r &= \rho \left(1 + \frac{M}{2 \rho} \right)^2 , \label{eq:rrho} \\
\Leftrightarrow \quad \rho &= \frac{1}{2} \left( r - M + \sqrt{(r-M)^2 - M^2} \right) . \label{eq:rhoSchw}
\end{align}
Here $\alpha(\rho)$ is a \emph{conformal factor}, and $n(\rho)$ is the effective \emph{refractive index} of the spacetime. 

 \subsubsection{The Schwarzschild interior}
The Schwarzschild \emph{interior} solution of Einstein's equations is sourced by the energy-momentum tensor of a body of constant density $\sigma$ with a radially-varying pressure $p(r)$, such that $p(R) = 0$ at the star's surface. In standard (Schwarzschild) coordinates, the interior Schwarzschild spacetime has a line element of the form (\ref{SSSS}) with 
\begin{equation}\label{intSfunctions}
    A(r) = \f{1}{4}\left(3\sqrt{B(R)} - \sqrt{B(r)}\right)^2~\quad \text{ and }\quad~B(r) = 1 - \f{2Mr^2}{R^3}~.
\end{equation}
The geometry is well-defined for stars of radius $R$ greater than the Buchdahl radius of $R_{\text{Buch}} \equiv 9 M / 4$. In the limit $R \rightarrow R_{\text{Buch}}$, a curvature singularity arises in the centre of the star.

This geometry can also be re-expressed in isotropic coordinates. To transform the line element \eqref{SSSS} into the form \eqref{isotropic-form}, it follows that we must have
\begin{align}
n(\rho) d\rho &= \frac{dr}{\sqrt{A(r) B(r)}}, &
n(\rho) \rho &= \frac{r}{\sqrt{A(r)}}. \label{eq:n-def}
\end{align}
After taking a ratio and integrating, one obtains
\begin{align} \label{eq:rho-integral}
\rho &= \exp\left( \int \frac{dr}{r \sqrt{B(r)}} \right) .
\end{align}
In certain cases, the integral can be obtained in closed form, leading to expressions for $\rho(r)$, $n(r)$ and, if the inversion is straightforward, $r(\rho)$ and $n(\rho)$. The Schwarzschild star is one such case. 

By integrating Eq.~\eqref{eq:rho-integral}, the isotropic coordinate is
\begin{equation}
    \rho = \frac{\rhoconst}{2} \cdot \frac{r}{1 + \sqrt{B(r)}}.  \label{eq:rho-interior}
\end{equation}
where $\rhoconst$ arises from the constant of integration. This can be inverted to obtain
\begin{equation}
    r = \frac{4 \rho}{\rhoconst (1 + \rho^2 / \beta^2)}~,  \label{eq:r-interior}
\end{equation}
where $\beta \equiv \rhoconst \sqrt{R^3/8M}$. To determine the constant $\rhoconst$, one can impose that $\rho$ is continuous at the surface of the star (i.e.~that the isotropic coordinate in the exterior matches the isotropic coordinate in the interior at $r=R$). Inserting $B(R) = 1-2M/R$, Eq.~\eqref{eq:rhoSchw} and $r=R$ into Eq.~\eqref{eq:rho-interior} yields
\begin{equation}
\rhoconst = \frac{1}{2} \left( 1 + \sqrt{1 - 2M/R} \right)^3.
\end{equation}
Furthermore, the functions $A(r)$ and $B(r)$ in Eq.~\eqref{intSfunctions} can be re-expressed in terms of the isotropic coordinate $\rho$ as
\begin{align}
\sqrt{A(\rho)} &= \frac{1}{2} \left(3 \sqrt{B(R)} - 1 \right) \cdot \frac{ 1 + \rho^2 / \rholens^2 }{1 + \rho^2 / \beta^2} , \label{eq:sqrtA}  \\
\sqrt{B(\rho)} &= \frac{1 - \rho^2 / \beta^2}{1 + \rho^2 / \beta^2} , \label{eq:sqrtB}
\end{align}
where  
\begin{equation}
\rholens \equiv \beta \cdot \frac{3 \sqrt{B(R)} - 1}{\sqrt{9B(R) - 1}} = R \cdot \frac{\gamma}{1 + 3 \sqrt{B(R)}} \sqrt{\frac{R-R_{\text{Buch}}}{M}}.
\end{equation}
Hence, by Eq.~\eqref{eq:n-def}, the effective refractive index is $n(\rho) = \tilde{\cal{R}} \hat{n}(\rho)$ where
\begin{align}
 \hat{n}(\rho) &= \frac{2}{1 + \rho^2 / \rho_{\text{lens}}^2} , \label{eq:n-fisheye} \\
 \tilde{\cal{R}} &= \frac{4}{\rhoconst \left(3 \sqrt{B(R)} - 1 \right)}. 
\end{align}
In summary, we have arranged the line element into the isotropic form of Eq.~\eqref{isotropic-form}, with $\alpha(\rho) = \sqrt{A(\rho)}$ given by Eq.~\eqref{eq:sqrtA}. It is straightforward to check that the refractive index of the exterior and interior match at $r=R$, as expected.

 \subsubsection{The Maxwell fisheye lens and the three-sphere}
The refractive index $\hat{n}(\rho)$ in Eq.~\eqref{eq:n-fisheye} is precisely that of a \emph{Maxwell fisheye lens}: a spherical lens that perfectly focusses the rays emanating from a point source placed on the rim of the lens at $\rho = \rho_{\text{lens}}$ at the antipodal point on the rim \cite{maxwell1890scientific}. This perfect-focussing property arises because the spatial part of the line element is conformal to a space of constant positive curvature: a three-sphere. This property can be made explicit through applying the coordinate transformation
\begin{equation}
    \rho = \rholens \tan( \chi / 2 ) ,
\end{equation}
where $\chi$ is a new coordinate representing an angle on a unit three-sphere. The line element becomes
\begin{align}\label{confIS}
 ds^2 = \widehat{\alpha}^2(\chi) \left( - dT^2 + d\chi^2 + \sin^2 \chi d\Omega^2 \right) , 
\end{align}
where $\widehat{\alpha}^2(\chi) = \lambda A(\chi)$, $dT = dt / \sqrt{\lambda}$ and 
\begin{align}
 \sqrt{A(\chi)}   &=  \frac{\left( 3 \sqrt{1 - 2M/R} - 1 \right)}{2 \left(\cos^2 (\chi / 2) + \mu^2 \sin^2(\chi/2) \right)} , \\
 &= \frac{1}{2} \left( 3 \sqrt{1 - 2M/R} - \frac{1 - \mu^2 \tan^2 \chi / 2}{1 + \mu^2 \tan^2 \chi / 2} \right), \\
 \lambda &\equiv \tilde{\cal{R}}^2 \rholens^2 = \frac{R^4}{4M (R - R_{\text{Buch}})} , 
\end{align}
and $\mu \equiv \rholens / \beta$. 

In summary, the Schwarzschild interior spacetime is conformal to a region ($0 \le \chi \le \chi_0$) of the spacetime of a unit three-sphere. On the surface of the star, $r=R$, the corresponding angle on the three-sphere $\chi_0$ is given by
\begin{equation} \label{eq:coschi0}
 \chi_0 = \cos^{-1} \left( \frac{R-3M}{\sqrt{R(R-2M)}} \right) 
 = \sin^{-1} \left(\sqrt{\frac{4 M (R - R_{\text{Buch}})}{R (R-2M)}} \right).
\end{equation}
Hence a compact star of radius equal to the light-ring radius, $R=3M$, encompasses exactly \emph{half} of the conformal sphere ($\chi_0 = \pi/2$); and as $R$ approaches the Buchdahl bound, the whole sphere is captured (i.e.~$\chi_0 \rightarrow \pi$).

 \subsubsection{Mode solutions}

To compute the self-force \emph{outside} a star via Eq.~\eqref{starSF2} one needs structure factors $S_\ell$ which in turn depend on the coefficients $\eta_\ell$ defined in Eq.~\eqref{eq:eta} in terms of the interior solution $\varphi^{\text{int}}_\ell$. Moreover, to compute the SF \emph{inside} the star one requires \emph{two} linearly-independent solutions in the interior. Here we obtain these solutions in a closed form.

To solve the wave equation in the interior spacetime we work with the metric  (\ref{confIS}) and make use of the conformal symmetry to first find solutions on a geometry $\widetilde{ds}^2$ with $ds^2 = \widehat{\alpha}^2(\chi) \widetilde{ds}^2$. Since we work with a conformally-coupled scalar field, the structure of the dynamical equation remains the same in the new (conformal) manifold as well \cite{Birrell_Davies_1982} given by, 
\begin{equation}
    \left(\widetilde{\Box} - \xi \widetilde{R}\right)\widetilde{\varphi} = 0~,
\end{equation}
for the conformal coupling $\xi = 1/6$. Here the tilded variables are objects in the conformally-related spacetime. In four spacetime dimensions, we have the relation $\varphi = \widetilde{\varphi} / \widehat{\alpha}$, that is,
\begin{equation}\label{transphi}
    \widetilde{\varphi} = \sqrt{\lambda A(\chi)} \, \varphi~.
\end{equation}
In the conformal three-sphere geometry, the Ricci scalar is simply $\widetilde{R} = 6$. Separating variables as usual and using the spherical harmonics, one obtains the following ``radial" equation in the (conformally) transformed variable $\chi$, 
\begin{equation}\label{conformalchi}
    \dv{}{\chi}\left(\sin^2{\chi}\dv{\widetilde{\varphi}_\ell}{\chi}\right) - \left(\sin^2{\chi}+ l (l+1)\right)\widetilde{\varphi}_\ell(\chi) = 0~.
\end{equation}
A pair of (homogeneous) solutions on the \emph{physical} spacetime are 
\begin{align}
    \hat{p}_{\ell}(r) = \f{P_{-1/2}^{-l-1/2}\left(\cos{\chi}\right)}{\sqrt{\lambda A(r)\sin{\chi}}}~, & \quad ~\hat{q}_{\ell}(r) = \f{P_{-1/2}^{l+1/2}\left(\cos{\chi}\right)}{\sqrt{\lambda A(r)\sin{\chi}}}~, \label{eq:pqint}
\end{align}
with $\chi$ related to $r$ by
\begin{align} \label{eq:rchi}
    \tan{\chi/2} &= \f{1}{\mu}\sqrt{\f{R^3}{2M}}\f{1}{r}\left[1 - \sqrt{B(r)}\right]~.
\end{align}
and inversely,
\begin{align}
  r &= R \frac{2 \sin \chi}{(3 \sqrt{B(R)} + \cos \chi)} \sqrt{\frac{R - R_{\text{Buch}}}{M}}
\end{align}
The $\hat{p}_{\ell}(r)$ function is regular at the origin ($r=0$), whereas the $\hat{q}_{\ell}(r)$ function is not.

If the charge is placed outside the star, we need only \emph{one} of the above solutions to compute the self-force. For the analysis in the previous section, we use 
\begin{equation}\label{phiint}
    \varphi_\ell^{\text{int}} = \hat{p}_{\ell}(r) = \left(\frac{R^4}{4 M (R-  R_\text{Buch})}\right)^{-1/2}\frac{ P_{-1/2}^{-l-1/2}(\cos (\chi ))}{\sqrt{\sin (\chi )} \left(\f{1}{2}\f{1-\mu^2\tan^2{\chi/2}}{1+\mu^2\tan^2{\chi/2}} - \f{3}{2}\sqrt{1-\f{2M}{R}}\right)}~.
\end{equation}
This leads to a closed-form expression for $\eta_{\ell}$ (defined in
Eq.~\eqref{eq:eta}), viz.,
\begin{align}
    \eta_{\ell} &= 
     -1 + \frac{(\ell+1)}{\sqrt{1-2M/R}} 
     \frac{P_{+1/2}^{-\ell-1/2}(\cos\chi_0)}{P_{-1/2}^{-\ell-1/2}(\cos\chi_0)}~,
\end{align}
where $\cos \chi_0$ was defined in Eq.~\eqref{eq:coschi0}. This can then be used to compute the structure factor in Eq.~\eqref{eq:structure-fac} and the self-force difference in Eq.~\eqref{starSF2}.

\subsubsection{Interior self force}\label{scalarisf}
When the charge is placed inside the star, we need both solutions in Eq.~\eqref{eq:pqint} to perform a matching procedure as in Sec.~\ref{sec:differencereg}.
The \textit{physical} field sourced by a particle at $r=r_0$ (see \eqref{genscalartortoise}) has mode solutions
\begin{equation}\label{hgsoln}
    u_{\ell}(r) = r\begin{cases}
        \Lambda_\ell\f{\hat{p}_{\ell}(r)}{\hat{p}_{\ell}(r_0)} & r \leq r_0 \\
        \kappa_\ell \f{\hat{p}_{\ell}(r)}{\hat{p}_{\ell}(r_0)} + \epsilon_\ell \f{\hat{q}_{\ell}(r)}{\hat{q}_{\ell}(r_0)} & r \geq r_0~,
    \end{cases}
\end{equation}
The matching procedure yields $\Lambda_\ell = \kappa_\ell + \epsilon_\ell$ with
\begin{align}\label{ISFdef}
    \kappa_{\ell} &= -\epsilon_\ell \sigma_{\ell}\f{\hat{p}_{\ell}(r_0)}{\hat{q}_{\ell}(r_0)}~, & 
    \epsilon_{\ell} &= -\f{4\pi}{\sqrt{B_0}}\sqrt{\f{2l+1}{4\pi}}\f{q}{r_0^2}\f{\hat{p}_{\ell}(r_0)\hat{q}_{\ell}(r_0)} {\mc{W}_{\ell}(r_0)}~, \\    
    \sigma_{\ell} &= \f{R\hat{q}'_{\ell}(R)- \tau_{\ell}~\hat{q}_{\ell}(R)}{R\hat{p}'_{\ell}(R)- \tau_{\ell}~\hat{p}_{\ell}(R)}~,
 & \quad \tau_{\ell} &= \dv{\log{Q_{\ell}(z)}}{\log{r}}.
\end{align}
The `bare' averaged modes of the SF can then be computed as a mode-sum as before: 
\begin{equation}\label{intSF}
    F_r^{\ell, \text{bare}} = - \f{1}{2}\left(\f{q}{r_0}\right)^2\f{2\ell+1}{\sqrt{B_0}\mc{W}_{\ell}}\left\{\hat{p}'_{\ell}\hat{q}_{\ell} + \hat{q}'_{\ell}\hat{p}_{\ell} - 2\sigma_{\ell}\hat{p}'_{\ell}\hat{p}_{\ell}\right\}~,
\end{equation}
where it is understood that the Wronskian is taken with respect to $r$.

\subsubsection{Difference regularization in the interior\label{sec:diff-interior}}
By analogy with Eq~(\ref{starSF2}), the mode sum for the (regularized) self-force can be split into two parts, $F_r = F_r^{(0)} + \Delta F_r$, where
\begin{align}
F_r^{(0)} &= q^2 \sum_{\ell=0}^\infty \left\{ -\frac{1}{2r_0^2}  \frac{2 \ell + 1}{\sqrt{B_0} \mathcal{W}_\ell} \left( \hat{p}'_{\ell}\hat{q}_{\ell} + \hat{q}'_{\ell}\hat{p}_{\ell} \right) - \widetilde{B}_\ell \right\}~, \label{eq:Fr0interior} \\
\Delta F_r &= \frac{q^2}{r_0^2} \sum_{\ell=0}^\infty \frac{2 \ell + 1}{\sqrt{B_0} \mathcal{W}_\ell} \sigma_\ell \hat{p}'_{\ell}\hat{p}_{\ell}~, \label{eq:Frdiff-interior}
\end{align}
where $\widetilde{B}$ is the regularization parameter in Eq.~\eqref{scalarregparam}. 
Remarkably, the first sum $F_r^{(0)}$ is found to vanish and hence the self-force can be calculated from the difference piece $\Delta F_r$ only; we return to this point in Sec.~\ref{sec:identities}. Hence $\sigma_\ell$ plays the role of a structure factor for the Schwarzschild interior.

\subsection{Electromagnetic self force} \label{sec:EMfield}
Overall, the analysis of the static EM self-force closely follows that of the  scalar field \cite{Isoyama_2012}. We comment here on the main differences in the analysis and in the results. 
The dynamical equation for the EM field is given by
\begin{equation}
    \nabla_\nu \mc{F}^{\nu \mu} = -4\pi j^\mu = -4\pi q\int_c u^\mu~\delta^4(x,z(\tau))~d\tau~.
\end{equation}
For static EM fields, the gauge potential (in Lorenz gauge) takes the form $\mc{A}_\mu = \left(\varphi(\textbf{x}),0,0,0\right)$. Of the four Maxwell equations, only the temporal equation is nontrivial. One performs a similar mode-sum decomposition as in (\ref{fielddecomp}) and (\ref{DiracYLM}) for the field and the source respectively, to obtain the radial differential equation, 
\begin{equation*}
    \dv[2]{u_{\ell}}{r_*} - V(r)u_{\ell} = 4\pi q \sqrt{\f{2l+1}{4\pi}}\f{\delta(r_*-r_{*0})}{r_0}~,
\end{equation*}
where 
\begin{equation*}
    V(r) =  \f{B}{A} \left[\f{l(l+1)}{r^2B} + \f{1}{2 r}\left(\f{AB'-A'B}{AB}\right)\right]~.
\end{equation*}
From here on, we proceed as in the scalar case to solve for the homogeneous solutions to the above ODE and use them to build an ansatz for the EM potential. In the exterior Schwarzschild spacetime the mode functions are given by, 
\begin{align}\label{EMhg}
    \hat{p}_{\ell} &= \begin{cases}
        1~, & l = 0~, \\
        (r-2M) P_{\ell}'(z)~, & l \neq 0~,
    \end{cases} \\ 
    \hat{q}_{\ell} &= (r-2M)Q_{\ell}'(z)~,
\end{align}
with $z = r/M-1$ as before, and ${}^\prime$ denoting a derivative with respect to $r$, 
such that the Wronskian of the two solutions is simply $\mc{W}_\ell = \ell (\ell + 1) M / r^2$ for $\ell > 0$ and $\mc{W}_\ell = M / r^2$ for $\ell = 0$.

To compute the mode functions in the interior, we use the fact that Maxwell's equations are conformally invariant \cite{Birrell_Davies_1982}. That is, if $\widetilde{F}_{\mu \nu}$ is a solution to Maxwell's equations on the (conformal) spacetime $d\widetilde{s}^2$ then $F_{\mu \nu} \equiv \widetilde{F}_{\mu \nu}$ is a solution on the (physical) spacetime $ds^2 = \hat{\alpha}^2(x) d\widetilde{s}^2$. Thus, we can use the solutions of the Maxwell equations on the three-sphere geometry, i.e., using the line element inside the parentheses in (\ref{confIS}). This approach yields the mode functions
\begin{align}\label{EMinthg}
    \hat{m}_{\ell}(r) = \f{P_{1/2}^{-l-1/2}\left(\cos{\chi}\right)}{\sqrt{\sin{\chi}}}~&, \quad \quad~\hat{n}_{\ell}(r) = \f{P_{1/2}^{l+1/2}\left(\cos{\chi}\right)}{\sqrt{\sin{\chi}}},
\end{align}
where the relation between $r$ and $\chi$ is stated in Eq.~\eqref{eq:rchi}.

\subsubsection{Exterior self force}
The `bare' SF is then defined as $F_\mu = q \mc{F}_{\mu \nu} \dot{x}^\nu$, where $\dot{x}^\nu$ denotes the particle's four-velocity. In the exterior, the `bare' averaged modes of the radial component are given by
\begin{equation}
    F_r^{\ell, \text{bare}} = \f{1}{2}\left(\f{q}{r_0}\right)^2 \f{2\ell+1}{\mc{W}_\ell \sqrt{B_0}}\left\{\hat{p}_\ell\hat{q}'_\ell+\hat{q}_\ell\hat{p}'_\ell - 2 \mc{S}_\ell\hat{q}_\ell\hat{q}'_\ell\right\}~,
\end{equation}
where the EM structure factor is \cite{Isoyama_2012}
\begin{equation}\label{EMstr}
    \mc{S}_{\ell}(Z) = \f{\ell(\ell+1)P_{\ell}(Z)-(1+\eta_{\ell})(Z-1)P_{\ell}'(Z)}{\ell(\ell+1)Q_{\ell}(Z)-(1+\eta_{\ell})(Z-1)Q_{\ell}'(Z)}.
\end{equation}
The logarithmic derivative is $\eta_{\ell} \equiv \left. d \ln \hat{m}_\ell / d \ln r \right|_{r=R}$ and has the closed form expression
\begin{equation}
    \eta_{\ell} = 2\left(\f{R-3M}{R-2M}\right) + \f{\ell+2}{\sqrt{1-2M/R}}\f{P_{3/2}^{-\ell-1/2}\left(\cos{\chi_0}\right)}{P_{1/2}^{-\ell-1/2}\left(\cos{\chi_0}\right)}~,
\end{equation}
where $\cos{\chi_0}$ is given in Eq. \eqref{eq:coschi0}. 

\subsubsection{Interior self force}
The calculation of the interior EM SF closely follows the analysis in Sec. \ref{scalarisf}. Here, we present only the results. The averaged `bare' modes before regularisation are given by a mode-sum,
\begin{equation}\label{EMintSF}
    F_r^{\ell, \text{bare}} = \f{1}{2}\left(\f{q}{r_0}\right)^2 \f{2\ell+1}{\sqrt{B_0}\mc{W}_{\ell}} \left\{\hat{m}'_{\ell}\hat{n}_{\ell} + \hat{n}'_{\ell}\hat{m}_{\ell} - 2\Gamma_{\ell}~\hat{m}'_{\ell}\hat{m}_{\ell}\right\}~,
\end{equation}
where $\Gamma_{\ell}$ is the structure factor analogous to $\sigma_{\ell}$ (\ref{ISFdef}) in the scalar case, given by
\begin{align}\label{EMISFdef}
    \Gamma_\ell &= \f{R\hat{n}'_{\ell}(R)- \varrho_{\ell}~\hat{n}_{\ell}(R)}{R\hat{m}'_{\ell}(R)- \varrho_{\ell}~\hat{m}_{\ell}(R)} \\
    \text{with }~\varrho_\ell &= \dv{\log{\hat{q}_\ell}}{\log{r}}\bigg\vert_{r=R}~ = -1 + \f{l(l+1)}{Z+1}\f{Q_{\ell}(Z)}{Q_{\ell}'(Z)}~.
\end{align}

\subsubsection{Direct regularisation} 
The splitting of the (retarded) field into the $R$ and $S$ fields is a general framework for regularising SFs and applies to EM fields as well. In a similar fashion to that in the scalar case, one has the following regularisation parameters for an expansion of the EM $S$ field in the vicinity of the charge. For a static EM field, defining $e^{2\Psi} \equiv A(r)$, the regularisation parameters at a point $r = r_0 + \Delta$ (with $\Delta \rightarrow 0$) are given by \cite{PhysRevD.86.064033}
\begin{align}\label{EMregparam}
    \widetilde{A} = \f{1}{r^2}B^{-1/2}\text{sign}\left(\Delta\right), &\quad \widetilde{B} = \f{1}{2r^2}\left(1-r\Psi'\right), \quad \widetilde{C} = 0,  \\
    \widetilde{D} = \f{1}{16r^2}\left[\left(1-r\Psi'\right) - \right.&\left.\left(1-r\Psi' + 3r^2\Psi'^2 - r^3\Psi'^3 + 6r^2\Psi'' + 2r^3\Psi'''\right)B +\right. \nn \\
    &\left.\left(1-4r\Psi' -3r^2\Psi''\right)rB'+\left(1-r\Psi'\right)r^2B''\right]~.
\end{align}
(N.B.~The signs of these terms are opposite to those given in Ref.~\cite{PhysRevD.86.064033} due to our convention in Eq.~\eqref{eq:Fr-regularized}).

\subsubsection{Difference regularisation} \label{sec:differencereg}
The method of difference regularisation carries over to the EM field case in a straightforward way. However, the interpretation of the difference is now slightly modified. Firstly, unlike in the conformal scalar case, the monopole ($\ell = 0$) solutions for the radial ODE for the EM field in the BH and star spacetime are identical. Hence, when computing the difference the monopole terms cancel, and thus the series start with the dipole ($\ell = 1$). More importantly, the SF difference is not equal to the full SF because the electromagnetic self force for a static particle on Schwarzschild spacetime is \cite{Smith:1980tv} 
\begin{equation} \label{eq:FrBH}
F_r^{\text{BH}} = \frac{q^2 M}{r^3 \sqrt{1 - 2M/r_0}} . 
\end{equation}
The SF for the Schwarzschild star is $F_r = F_r^{\text{BH}} + \Delta F^{\text{star}}_r$, where
\begin{equation}\label{EMsfdiff}
   \Delta F^{\text{star}}_r = -  \left(\f{q}{M}\right)^2\sqrt{\f{z_0-1}{z_0+1}}\sum_{\ell=1}^{\infty}\left(2\ell+1\right) \mc{S}_{\ell} \left(Q_{\ell}(z_0) - \f{z_0-1}{\ell(\ell+1)}Q_{\ell}'(z_0) \right)  Q_{\ell}'(z_0)~.
\end{equation}
Again, the structure factor $\mc{S}_{\ell}$ in Eq. \eqref{EMstr} is a function of $R$, the radius of the star, but it does not depend on $r_0$, the position of the charge. All other functions in Eq.~\eqref{EMsfdiff} are evaluated at the position of the particle, and are independent of $R$.

Remarkably, in the interior region ($r_0 < R$), one can also calculate the self-force from a difference sum only. Starting with the bare modes in Eq.~\eqref{EMintSF}, and following the prescription in Sec.~\ref{sec:diff-interior}, one splits the regularized mode sum into two parts. The first part vanishes, as is shown in the next section.

\subsection{Mode sum identities\label{sec:identities}}
For static self-forces, it is apparent that some regularized mode sums turn out to be equal to zero, and other mode sums can be expressed in closed form in a simple way. In this section we give this phenomenon some further consideration.

\subsubsection{Exterior\label{sec:ident-ext}}
A first result is that the following regularized mode sum is zero:
\begin{equation}
\mathcal{S}_{(0)} \equiv \sum_{\ell = 0}^{\infty} \left\{ S_{(0)\ell}  - \gamma_{0}(z) \right\} = 0 , \quad \quad S_{(0)\ell} \equiv (2 \ell + 1) P_\ell(z) Q_\ell(z) ,
\end{equation}
where $\gamma_{0}(z) \equiv (z^2 - 1)^{-1/2}$. This result is proved in Appendix \ref{sec:proof} to follow as a consequence of Christoffel's formula.

Taking a derivative with respect to $z$, and moving the derivative inside the sum, we arrive at
\begin{equation}
\mathcal{S}_{(0)}^\prime = \sum_{\ell = 0}^{\infty} \left\{ (2 \ell + 1) \left( P_\ell'(z) Q_\ell(z) + P_\ell(z) Q_\ell'(z)  \right) - \gamma_0^\prime(z) \right\} = 0 . \label{eq:I0prime}
\end{equation}
By comparison with Eq.~(\ref{eq:FrBHsum}) and (\ref{eq:bell-tilde}), we conclude that the self-force on a static charge outside a black hole is precisely equal to zero. This is the result of Wiseman \cite{Wiseman:2000rm}, but here derived directly from the mode sum expression. 

Turning to the electromagnetic case in the Schwarzschild exterior, we can use the result above to prove the next identity, which is:
\begin{equation}
\mathcal{S}_{(1)} \equiv
\sum_{\ell = 1}^\infty \left\{ S_{(1)\ell} - \gamma_{1}(z) \right\} = \sigma_1(z), 
\end{equation}
where
\begin{equation} \label{eq:S1}
S_{(1)\ell} \equiv
\begin{cases}
\frac{2 \ell + 1}{\ell (\ell+1)} \left\{ (z-1) P_{\ell}^\prime(z) \right\} \left\{ (z-1) Q_{\ell}^\prime(z) \right\},  & \ell > 0, \\
(z - 1) Q'_{0}(z), & \ell = 0 ,
\end{cases} 
\end{equation}
and $\gamma_{1}(z)$ and $\sigma_1(z)$ functions to be derived below. To establish this, first we observe that
\begin{equation}
\frac{1}{(z^2-1)} \frac{d}{dz} \left((z+1)^2 S_{(1)\ell} \right) = (2 \ell + 1) \left\{P_\ell^\prime(z) Q_\ell(z) + P_\ell(z) Q_\ell^\prime(z) \right\} ,  
\end{equation}
where here we have used the Legendre differential equation,
\begin{equation}
\frac{d}{dz}\left((z^2 - 1) P_\ell(z)\right) = \ell (\ell + 1) P_{\ell}(z).
\end{equation}
Next, subtract $\gamma_0'(z)$, take the sum, and apply Eq.~(\ref{eq:I0prime}) to obtain
\begin{equation}
  \sum_{\ell=1}^{\infty} \left\{ \frac{d}{dz} \left( (z+1)^2 S_{(1)\ell} \right) - (z^2 - 1) \gamma_0^\prime(z) \right\} = (z^2 - 1) \left( \gamma_0^\prime - S_{(1)0}^\prime \right) ,
\end{equation}
with the $\ell = 0$ term moved across to the right-hand side. Now we can integrate and rearrange to obtain Eq.~\eqref{eq:S1} with
\begin{align}
\gamma_{1}(z) &= \frac{1}{(z+1)^2} \int (z^2 - 1) \gamma_{0}^\prime dz = - \frac{\sqrt{z^2 - 1}}{(z+1)^2} \\
\sigma_{1}(z) &= \gamma_{1} - \frac{1}{(z+1)^2} \int (z^2 - 1) S_{(1)0}^\prime(z) dz = \frac{z - \sqrt{z^2 - 1}}{(z+1)^2} .
\end{align}
Here the integration constants are justified by considering asymptotic behaviour.

\subsubsection{Interior\label{sec:ident-int}}
In the interior, for the EM case we can make use of the result that another regularized mode sum vanishes: $\mathcal{S}_{(1)}^{\text{int.}} = 0$, where
\begin{equation}
\mathcal{S}_{(1)}^{\text{int.}} =
\sum_{\ell = 0}^{\infty} \left\{ \mathcal{S}_{(1)\ell}^{\text{int.}} - \gamma_1^{\text{int.}} \right\} ,
\end{equation}
and
\begin{equation}
\mathcal{S}_{(1)\ell}^{\text{int.}} \equiv (2 \ell + 1) (-1)^\ell P_{1/2}^{+(\ell+1/2)}(x) P_{1/2}^{-(\ell + 1/2)}(x), \label{eq:S1int}
\end{equation}
and $\gamma_1^{\text{int.}} \equiv 2 / \pi$. This is shown by  first establishing the following result for the partial sum:
\begin{align}
\sum_{\ell = 0}^{n} \mathcal{S}_{(1)\ell}^{\text{int.}}  &= -\frac{x}{\pi} + (1-x^2)^{3/2} (-1)^{n+1} \left\{ P_{1/2}^{-N}(x) \partial_x P_{1/2}^{N+1}(x) - n(n+2) P_{1/2}^{-(N+1)}(x) \partial_x P_{1/2}^N(x) \right\} , \label{eq:interior-sum}
\end{align}
where $N \equiv n+1/2$. In Appendix \ref{sec:proof2} this is shown to follow as a consequence of the recurrence relation DLMF 14.10.1 \cite{NIST:DLMF}. Next, by considering the large-$n$ asymptotics of the right-hand side, one extracts the regularization function $\gamma_1^{\text{int}}$ and the limit of the infinite sum (zero).

A similar result holds in the scalar-field case, since one can show by the same methods that a partial sum is
\begin{align}
\sum_{\ell = 0}^{n} \mathcal{S}_{(0)\ell}^{\text{int.}}(x)  &= -\frac{x}{\pi} + (1-x^2)^{3/2} (-1)^{n+1} \left\{ P_{-1/2}^{-N}(x) \partial_x P_{-1/2}^{N+1}(x) - (n+1)^2 P_{-1/2}^{-N-1}(x) \partial_x P_{-1/2}^N(x) \right\} \label{eq:interior-sum2}
\end{align}
where 
\begin{equation}
\mathcal{S}_{(0)\ell}^{\text{int.}}(x)  \equiv (2 \ell + 1) (-1)^\ell P_{-1/2}^{-(\ell+1/2)}(x) P_{-1/2}^{\ell+1/2}(x) .
\end{equation}
By considering large-$n$ asymptotics of the right-hand side, it follows that
\begin{equation}
\mathcal{S}_{(0)}^{\text{int.}} \equiv \sum_{\ell = 0}^{\infty} \left\{ \mathcal{S}_{(0) \ell}^{\text{int.}}  - \gamma^{\text{int.}}_{0}(z) \right\} = 0 ,
\end{equation}
where $\gamma_0^{\text{int.}} \equiv 2 / \pi$. 
These results, along with their derivatives, show why the interior self-force can be found by evaluating the difference terms only.

\subsubsection{Numerics}

\begin{figure}
    \centering
    \includegraphics{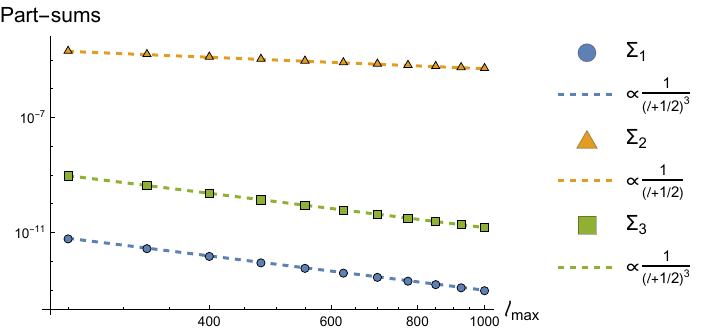}
    \caption{This figure shows the behaviour of the partial mode-sum as we increase the value of $\ell$ where we truncate the sum. The data $\Sigma_i$, $i=1,2,3$ denote the BH scalar sum \eqref{eq:FrBHsum}, the scalar interior sum \eqref{eq:Fr0interior} and the EM interior mode-sum respectively. In all cases, we see that the partial sums decrease according to a power law as we truncate the mode-sums at higher values of $\ell_{\text{max}}$, implying that the infinite sum must vanish, in accordance with the results in Secs.~\ref{sec:ident-ext} and \ref{sec:ident-int}.}
    \label{fig:part-sums}
\end{figure}

To support the arguments above, we have also evaluated the regularized mode sums numerically. Figure \ref{fig:part-sums} demonstrates that the partial sums converge towards zero as $n \rightarrow \infty$.

\section{Results\label{sec:results}}
In this section, we obtain new results for the self-force in the vicinity of a Schwarzschild star (i.e.~a matter distribution of constant density) in three forms. First, as a series expansion in the far-field ($r \gg R$) and near the centre of the star; second, as a leading-order approximation for the self-force near the surface of the star; and, third, as numerical data across the whole domain. As a first step, we examine the $\ell$-modes of the self-force and we verify that the two regularization approaches produce consistent results.

\subsection{Regularization and validation\label{sec:regforces}}
 
The bare modes of the SF can be straightforwardly computed, for a particle in the exterior or the interior of the star, by using Eq.~\eqref{starSF1} and Eq.~\eqref{intSF}, respectively. The SF difference in the exterior is given in Eq.~\eqref{starSF2}. As described in Sections \ref{directmethod}, \ref{diffmethod} and \ref{sec:identities}, we have the option of regularising with two different methods.  

\subsubsection{Exterior scalar self force}
\begin{figure}[htp]
\subfloat[Regularising scalar SF inside the Schwarzschild star]{%
  \includegraphics[width = 0.425\textwidth]{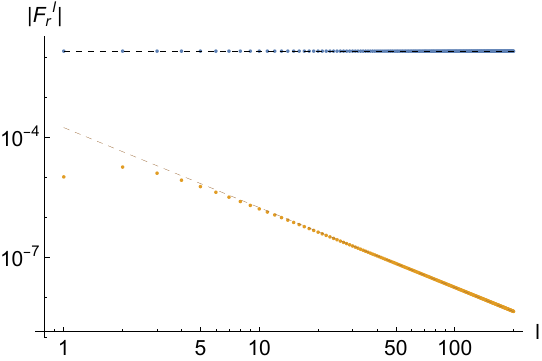}%
\label{intregcurves}}
\subfloat[Regularising scalar SF outside the Schwarzschild star]{%
  \includegraphics[width = 0.575\textwidth]{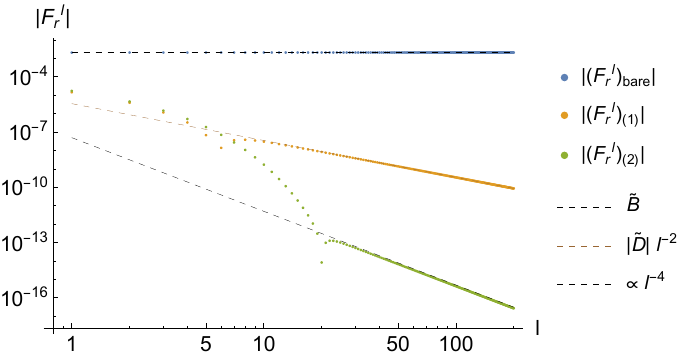}%
\label{extregcurves}}
\caption{Modes of the self force $F_r^\ell$ on a log-log scale showing the expected polynomial decay after subtracting regularization terms in the interior (left) and the exterior (right).
For the above plots we choose a star of radius $R = 11 M$ with the charge placed at $r_0 = 6 M$ (left) and $r_0 = 16M$ (right) in \ref{intregcurves} and \ref{extregcurves} respectively.
In all figures units such that $M=q=1$ are adopted.
}\label{verifying direct regularisation}
\end{figure}

For the exterior Schwarzschild scalar SF using the direct method we obtain $\ell$-modes for the force as shown in Fig \ref{extregcurves}. To test the regularisation procedure, we make the definitions 
\begin{align}\label{regforce}
    \left(F_r\right)^\ell_{(1)} &\equiv  F_r^{\ell, \text{bare}} - \widetilde{B} , \\
    \left(F_r\right)^\ell_{(2)} &\equiv  F_r^{\ell, \text{bare}} - \widetilde{B} - \f{\widetilde{D}}{\left(\ell-1/2\right)\left(\ell+3/2\right)} .
\end{align}
We verified that the (averaged) `bare' modes asymptote to the constant $\widetilde{B}$ as $\ell \rightarrow \infty$; and the regularised modes $\left(F_r\right)^\ell_{(1)}$ and $\left(F_r\right)^\ell_{(2)}$ fall off as $(\ell + 1/2)^{-2}$ and $(\ell + 1/2)^{-4}$, respectively, in this limit. Subtracting only the $\widetilde{B}$ parameters from the `bare' averaged modes in Eq.~\eqref{eq:Fl-bare} is sufficient to ensure convergence of the series; however, subtracting additional parameters lead to faster convergence, in line with the faster decay of the modes, and greater numerical accuracy of the sum.

We also examined the $\ell$-modes of the SF \emph{difference}. We verified  that the $\ell$-modes in Eq.~\eqref{starSF1} fall off \emph{exponentially} with $\ell$, as previously noted in \cite{Isoyama_2012}. We elaborate on this exponential convergence in later sections (see Figs.~\ref{surfcompare} and \ref{verifying regularisation}).

To compute the \emph{total} self-force one must evaluate the mode sum numerically. As is standard, we sum modes up to a suitably-large $\ell_{\text{max}}$ value, which (at minimum) must lie in the regime in which the regularization parameters provide a good fit. For direct regularization, we then estimate the `large-$\ell$ tail' remainder by fitting the modes to a polynomial in powers of $(\ell+1/2)^{-4}$, and then computing the sum of this polynomial from $\ell_{\text{max}} + 1$ to infinity. For difference regularization, the `large-$\ell$ tail' is modelled as an exponentially-converging sum.

Numerical results for $F_r$ obtained via the two regularization methods are given in Table \ref{tbl:sample-results}. The data shows agreement at least 6 decimal places close to the surface of the star, and the level of agreement increases with $r - R$. The accuracy can be further improved by modifying internal parameters at the expense of run-time (e.g.~by increasing $\ell_{\text{max}}$). It is likely that the results from the difference method are more accurate than the direct method. 

\begin{table}
  \begin{tabular}{| l | l | l |}
  \hline
 $r_0 / M$ & $F_r / (q/M)^2$ (scalar)  & $F_r / (q/M)^2$ (EM)  \\
 \hline
 $6.1$ & $5.200896(6) \times 10^{-3}$ &$1.604917(2) \times 10^{-2}$ \\
  & $5.2008963(3) \times 10^{-3}$ &$1.6049174(1) \times 10^{-2}$  \\ \hline
$6.5$  & $2.71508989(1) \times 10^{-3}$  &$8.67227486(1) \times 10^{-3}$  \\
   &$2.71508989852 \times 10^{-3}$ &$8.67227486152\times 10^{-3}$  \\ \hline
$7$  &$1.712197896(4) \times 10^{-3}$ & $5.64183966(1) \times 10^{-3}$ \\
  &$1.71219789621 \times 10^{-3}$ & $5.64183966893 \times 10^{-3}$  \\ \hline
$8$  &$8.93714041(2) \times 10^{-4}$ & $3.080427341(2) \times 10^{-3}$ \\
  &$8.93714041480 \times 10^{-4}$ & $3.08042734138 \times 10^{-3}$  \\ \hline
$10$ &$3.627255060(4) \times 10^{-4}$ &$1.3213072171(6) \times 10^{-3}$   \\
  &$3.62725506014 \times 10^{-4}$ & $1.32130721711 \times 10^{-3}$    \\ \hline
$15$  &$8.743329274(3) \times 10^{-5}$ & $3.3892349544(5) \times 10^{-4}$  \\
  &$8.74332927466 \times 10^{-5}$ & $3.38923495442 \times 10^{-4}$  \\ \hline
$25$  &$1.6863404917(1) \times 10^{-5}$ & $6.8103725730(2) \times 10^{-5}$  \\
  &$1.68634049176 \times 10^{-5}$ & $6.81037257301 \times 10^{-5}$  \\ \hline
  \end{tabular}
    \caption{Sample results for the self-force $F_r$ calculated by direct regularization (upper row), and by difference regularization (lower row), \emph{outside} a Schwarzschild star of radius $R = 6M$. The mode-sum is truncated at $\ell_{\text{max}}=200$ (with the remainder of the modes fitted by a power-law) in all cases except at $r_0/M = 6.1$, for which we use the heuristic $\ell_{\text{max}} = 2\pi R/\Delta r \sim 440$. For the EM case (right column), the BH SF is added to the difference calculation  so as to compare with the direct method. The numeral in parentheses shows an estimate of the numerical error in the last quoted digit; where absent, the result should be correct in all digits stated. This error estimate was derived from the `large-$\ell$ tail' fit. 
    }
    \label{tbl:sample-results}
\end{table}
 
\subsubsection{Interior scalar self force}
In the interior ($r_0 < R$) we regularise directly by using the parameters in Eqs.~\eqref{scalarregparam} and Eqs.~\eqref{EMregparam} (which are reproduced from Ref.~\cite{PhysRevD.86.064033}) with the metric functions $A(r)$ and $B(r)$ of the Schwarzschild interior solution. The $\ell$-modes of the SF before and after regularisation are shown in Fig.~\ref{intregcurves}. 

In the \emph{interior} we cannot apply the $\tilde{D}$ regularization terms from Ref.~\cite{PhysRevD.86.064033}, because this was calculated for a \emph{minimally-coupled} scalar field, and we are here considering a conformally-coupled field. Conversely, in the exterior the parameter $\tilde{D}$ can be applied, because the Ricci scalar vanishes in the vacuum region and consequently there is no difference in the (locally-defined) S field between the two cases.

\begin{table}
  \begin{tabular}{| l | l | l |}
  \hline
 $r_0 / M$ & $F_r / (q/M)^2$ (scalar)  & $F_r / (q/M)^2$ (EM)  \\
 \hline
 $0.5$ & $1.9534(6)\times 10^{-4}$ & $5.1807268554 \times 10^{-4}$ \\
  & $1.95342961682 \times 10^{-4}$ & $5.1807268554 \times 10^{-4}$ \\ \hline
$1$  & $3.9219(2) \times 10^{-4}$  & $1.0432023863 \times 10^{-3}$ \\
   &$3.92187625293 \times 10^{-4}$ &$1.0432023863 \times 10^{-3}$\\ \hline
$2$  &$7.9789(1)  \times 10^{-4}$ &$2.1477348199 \times 10^{-3}$  \\
  &$7.97886997930 \times 10^{-4}$ &  $2.1477348199 \times 10^{-3}$ \\ \hline
$3$  &$1.24040(6) \times 10^{-3}$ &$3.4067324914 \times 10^{-3}$  \\
  &$1.24040470355 \times 10^{-3}$ &$3.4067324914 \times 10^{-3}$ \\ \hline
$5$ &$2.60490(1) \times 10^{-3}$ & $7.642467(7) \times 10^{-3}$  \\
  &$2.60486(5) \times 10^{-3}$ &  $7.64248884062 \times 10^{-3}$   \\ \hline
$5.9$  &$5.482220(7) \times 10^{-3}$ & $1.667817961(1)\times 10^{-2}$  \\
  &$5.4822155(1) \times 10^{-3}$ &$1.66781796376 \times 10^{-2}$ \\ \hline
  \end{tabular}
    \caption{Sample results for the self-force $F_r$ calculated by direct regularization (upper row), and by difference regularization (lower row), \emph{inside} a Schwarzschild star of radius $R = 6M$. The direct mode-sum is truncated at $\ell_{\text{max}}=1200$ ($\ell_{\text{max}}=3000$) for the scalar (EM) case, with the remainder of the modes fitted by a power-law. The difference mode-sum is truncated at $\ell_{\text{max} = 25}$ ($\ell_{\text{max} = 50}$ for the scalar (EM) case. We note that the scalar force is direct regularised with the $\widetilde{B}$ parameter only, while the EM force has both the $\widetilde{B}$ and the $\widetilde{D}$ parameters available. This explains the better agreement between the direct and the difference results for the EM case. The numeral in parentheses shows an estimate of the numerical error in the last quoted digit. This error estimate was derived from the `large-$\ell$ tail' fit. For the direct regularised scalar force, the error estimate is derived from $0.1\%$ of the fitted tail. 
    }
    \label{tbl:int-sample-results}
\end{table}

\subsection{Asymptotics and series expansions\label{sec:series}}
We obtain series expansions for the self-force by inserting asymptotic expansions of the mode functions into the mode sums for the self-force difference, such as Eq.~\eqref{eq:Frdiff}. The low-$\ell$ modes provide good approximations both near the star's centre, and in the far field.

\subsubsection{Scalar self force: far-field}

In the far-field region where $r \gg R$ and $r \gg M$, the scalar self-force has an asymptotic expansion
\begin{equation}\label{scalarexp}
\mathcal{F}_r = \frac{q^2 M}{3 r^3} \left( \left( 1 + \frac{2 M}{r} + \frac{23 M^2}{6 r^2} + \ldots \right) \widehat{S}_0\left( \zeta \right)  + \left(\f{ 2 R^2}{5 r^2} + \ldots \right) \widehat{S}_{1}\left( \zeta \right) + \ldots \right)   
\end{equation}
where $\widehat{S}_\ell(\zeta)$ denote normalised structure factors with $\zeta \equiv M/R$; for the monopole ($\ell = 0)$ and dipole ($\ell = 1$) terms we obtain
\begin{align}
 \widehat{S}_0(\zeta) &= 1 - \frac{8}{5} \zeta - \frac{67}{105} \zeta^2 -\f{368}{315} \zeta^3 -\f{3589}{1925} \zeta^4 -\f{2235496}{675675} \zeta^5 + \mc{O}(\zeta^{6}) , \label{eq:S0s0} \\
 \widehat{S}_1(\zeta) &= 1 - \frac{32}{21} \zeta + \frac{158}{315} \zeta^2 - \frac{9472}{7425} \zeta^3 -\frac{298996}{257985} \zeta^4 +\mc{O}\left(\zeta^5\right). \label{eq:S1s0}
\end{align}

\begin{figure}[ht]
\centering
\subfloat[Far field expansions vs numerics: Scalar SF]{%
\includegraphics[width=15cm]{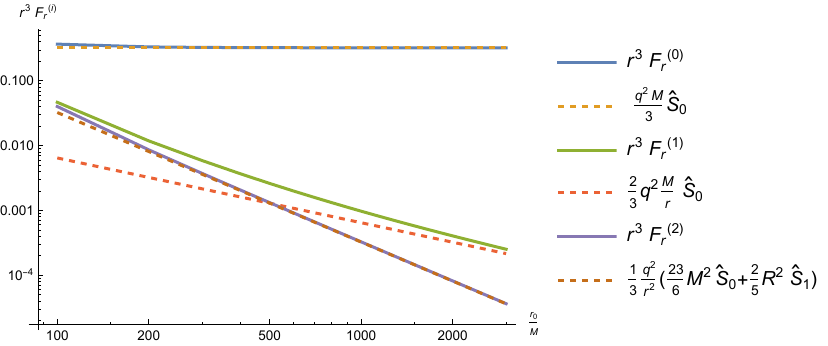}%
\label{Scomp}}%
\vspace{5 px}
\subfloat[Far field expansions vs numerics: Electromagnetic SF]{%
\includegraphics[width=15cm]{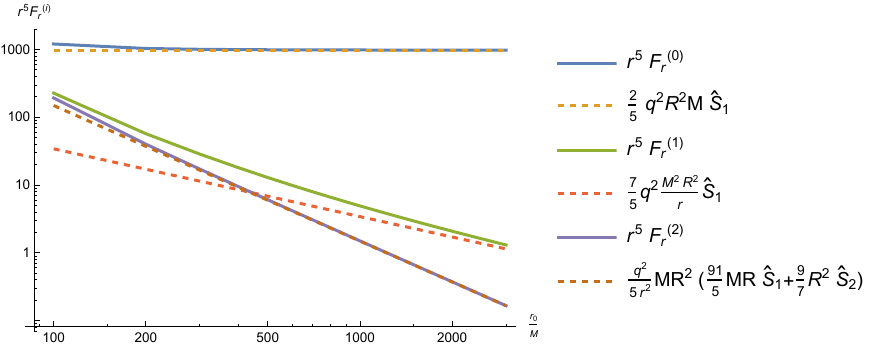}%
\label{EMcomp}}%
\caption{A comparison of the numerical SF data obtained by evaluating the mode-sum results [solid] for the scalar and electromagnetic SFs, with the series expansions [dashed] in Eq.~\eqref{scalarexp} and Eq.~\eqref{EMexp} respectively, for a star of radius $R = 50M$.}
\end{figure}

With expansions such as Eq.~\eqref{scalarexp}, one can check against numerical data order-by-order in $M/r$. We introduce the following notation where the calligraphic $\mc{F}^{(i)}$ represents the difference between the numerically-calculated force $F_r$ and the the series expansion in (\ref{scalarexp}) truncated at the $i$th order. That is, 
\begin{align}
    \mc{F}_r^{(0)} &= F_r~, \\
    \mc{F}_r^{(1)} &= \mc{F}_r^{(0)} - q^2\frac{M}{3 r^3} \widehat{S}_0 \left(\tfrac{M}{R} \right)~, \\
    \mc{F}_r^{(2)} &= \mc{F}_r^{(1)} - q^2\frac{2 M^2}{3 r^4} \widehat{S}_0 \left( \tfrac{M}{R} \right)~.
\end{align}
Figure \ref{Scomp} shows these quantities across a range of $r/M$, verifying the expected behaviour.

\subsubsection{Scalar self force: centre of the star}

One can obtain expansions near the centre of the star in a similar way.  To do so, we exploit the applicability of difference regularisation in the interior and use large-$R$ expansions for the mode functions. For $R\gg M$ and $r \ll R$, 
\begin{align}\label{eq:Fcentre-s0}
    F^\text{int}_r &= \f{q^2M r}{3 R^4}\left\{\widehat{\sigma}_1 -\f{M}{R}\widehat{\sigma}_0 + \f{r^2}{R^2}\left[\f{2}{5}\widehat{\sigma}_2+\f{M}{R}\left(\f{29}{10}\widehat{\sigma}_1+\f{3}{4}\widehat{\sigma}_2\right) + \f{M^2}{R^2}\left(\f{123}{64}\widehat{\sigma}_2+\f{45}{16}\widehat{\sigma}_1-\f{46}{15}\widehat{\sigma}_0\right)\right] + \right. \nn \\
    & \quad \quad \quad \quad \quad \left. + \mathcal{O}\left(\f{r^4}{R^4}\right) \right\}~,
\end{align}
where the $\widehat{\sigma}_{\ell} \equiv \widehat{\sigma}_{\ell}(\zeta)$ are (normalised) structure factors for the exterior defined in (\ref{ISFdef}) with $\zeta \equiv M/R$ as before. The force at the centre of the star is exactly zero, due to spherical symmetry, and the leading term in the force arises from the dipole rather than the monopole. The monopole, dipole and quadrupole structure factors have the series expansions
\begin{align}
    \widehat{\sigma}_0(\zeta) &= 1-\frac{4}{3}\zeta-\frac{31}{12}\zeta^2-\frac{298}{45}\zeta^3-\frac{122771}{7560}\zeta^4 - \frac{297683}{7560}\zeta^5 + \mc{O}(\zeta^6)\\
    \widehat{\sigma}_1(\zeta) &=1+\frac{79}{30}\zeta+\frac{151}{36}\zeta^2+\frac{89099}{9450}\zeta^3+\frac{6125213}{283500}\zeta^4+\frac{533398271}{10692000}\zeta^5 + \mc{O}(\zeta^6) \\
    \widehat{\sigma}_2(\zeta) &= 1-\frac{2553}{280}\zeta-\frac{109251}{22400}\zeta^2-\frac{882841}{18816000}\zeta^3-\frac{2780229721}{77271040000}\zeta^4+ \mc{O}(\zeta^5) .
\end{align}

It is notable in Fig.~\ref{fig:fulldomainSF} that the self-force on a particle held at $r=r_0$ near the centre of the star scales in linear proportion to $r_0$, at leading order. A similar linear scaling with $r_0$ was found in the case of (the interior of) a mass shell in Ref.~\cite{Burko_2000}. There is one important difference with the shell case, however. For the Schwarzschild star, the force is positive (i.e.~directed away from the centre) across the entire domain, whereas for the shell case it repels from the shell (i.e.~it changes sign at $r=R$). For the shell case, there is Simple Harmonic Motion around the centre at $r=0$; for the star, this is not the case, due to the sign difference.

\subsubsection{Electromagnetic self force: far-field}
For the EM case, we have a similar far-field series expansion with minor differences. In the EM case, the monopole structure factor vanishes identically, because the monopole contribution to the SF in the BH and the star spacetime coincide (and this would also be the case for a \emph{minimally coupled} scalar field). This is just a restatement of the fact that at the monopole level, the SF is universal between stellar and BH spacetimes (with a minimally coupled scalar) \cite{Drivas:2010wz}. Thus, we start with expansions of the dipole term in the SF difference mode sum in (\ref{EMsfdiff}). 
\begin{equation}\label{EMexp}
\Delta \mc{F}_r^{EM} =  \frac{ 2 q^2 M}{5}\f{R^2}{r^5}\left(
\left(1 + \frac{7 M}{ 2 r} + \f{91 M^2 }{10 r^2} + \ldots \right)
\widehat{\mc{S}}_1 (\tfrac{M}{R})
+
\left(\f{9 R^2}{14 r^2} + \ldots \right) \widehat{\mc{S}}_2 \left(\tfrac{M}{R} \right) + \ldots \right)~, 
\end{equation}
where the normalised dipole and quadrupole structure factors $\widehat{\mc{S}}_1 \left(\tfrac{M}{R} \right)$ and $\widehat{\mc{S}}_2 \left(\tfrac{M}{R} \right)$ have the expansions (with $\zeta = M/R$ as before): 
\begin{align}
 \widehat{\mc{S}}_1(\zeta) &= 1 - \frac{8 }{7}\zeta-\frac{61}{105}\zeta^2- \frac{472}{825}\zeta^3 - \frac{392507}{525525}\zeta^4 + \mc{O}\left(\zeta^5\right) ,  \label{eq:S1s1} \\
 \widehat{\mc{S}}_2(\zeta) &= 1 - \frac{184}{45}\zeta+\frac{10152}{1925} \zeta^2 - \frac{4496896 }{2627625}\zeta^3 - \frac{566696}{1299375}\zeta^4 +\mc{O}\left(\zeta^5\right). \label{eq:S2s1}
\end{align}

Once again, these expansions offer a way to compare the analytic expressions with our numerical data, so we define the notation (as before)
\begin{align}
    \left(\Delta \mc{F}_r^{EM}\right)^{(0)} &= \Delta F_r^{EM}~, \\
    \left(\Delta \mc{F}_r^{EM}\right)^{(1)} &= \left(\Delta \mc{F}_r^{EM}\right)^{(0)} - \frac{ 2 q^2}{5M^2}\f{R^2}{r^5}\widehat{\mc{S}}_1~, \\
    \left(\Delta \mc{F}_r^{EM}\right)^{(2)} &= \left(\Delta \mc{F}_r^{EM}\right)^{(1)} -\frac{ 7 q^2}{5M}\f{R^2}{r^6}\widehat{\mc{S}}_1~.
\end{align}
Figure \ref{EMcomp} compares the numerical data with these far field expansions, and we obtain similar agreement as in the scalar field case.

\subsubsection{Electromagnetic self force: centre of the star}
As before, we can look at series expansions near the centre of the star for the EM SF using large $R$ expansions and difference regularisation. However, the monopole contribution to the difference mode-sum is identically zero. This is because the monopole mode function in the interior $\hat{m}_0$ is a constant. Since the interior difference term in (\ref{EMintSF}) is $\propto \hat{m}'$, this term is identically zero. Thus, the series expansions only start with the dipole terms.
\begin{equation} \label{eq:Fcentre-s1}
    \left(\mc{F}_r^{EM}\right)^{\text{int}} = \f{q^2 M r}{R^4}\left[\widehat{\Gamma}_1 + \f{r^2}{R^2}\left(\f{2}{5}\widehat{\Gamma}_2 + \f{M}{R}\left(\f{23}{10}\widehat{\Gamma}_1 + \f{3}{4}\widehat{\Gamma}_2\right)+ \f{M^2}{R^2}\left(\f{39}{16}\widehat{\Gamma}_1 + \f{123}{64}\widehat{\Gamma}_2\right)\right) + \ldots\right]
\end{equation}
where the normalised structure factors $\widehat{\Gamma}_\ell$ have the expansions, 
\begin{align}
    \widehat{\Gamma}_1 &= 1 +\frac{13}{10}\zeta +\frac{11}{4}\zeta^2 +\frac{1089}{175}\zeta^3 +\frac{75193}{5250}\zeta^4 + \frac{4374767}{132000}\zeta^5 + \mc{O}(\zeta^6) \\
    \widehat{\Gamma}_2 &= 1 -\frac{1881}{280}\zeta -\frac{5379}{22400}\zeta^2 +\frac{1675847}{18816000}\zeta^3 +\frac{11981145639}{77271040000}\zeta^4 + \mc{O}(\zeta^5) .
\end{align}

\subsection{Approach to the boundary\label{sec:boundary}}

\subsubsection{Scalar self force}
\begin{figure}
\includegraphics[width = 0.75\linewidth]{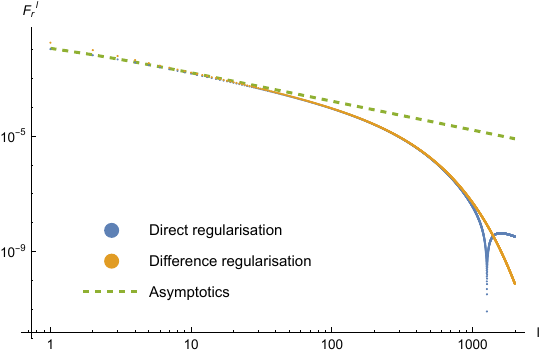}
\caption{Numerical data for the $\ell$-modes of the SF (dots)  on a log-log plot, 
 for a particle close to the star surface at $R=3M$, with $\Delta r = r_0 - R = 0.005 M$. The expected exponential ($(l+1/2)^{-2}$) behaviour for the difference (direct) regularised modes only starts after an initial regime with a slower decay. The slow-decay behaviour is compared with the green dashed guideline, which is proportional to $(l+1/2)^{-1}$ and with a coefficient taken from Eq.~\eqref{surfexp}. }
\label{surfcompare}
\end{figure}

\begin{figure}[htp]
\subfloat[Direct regularisation near the star boundary from the exterior]{%
  \includegraphics[width = 0.5\textwidth]{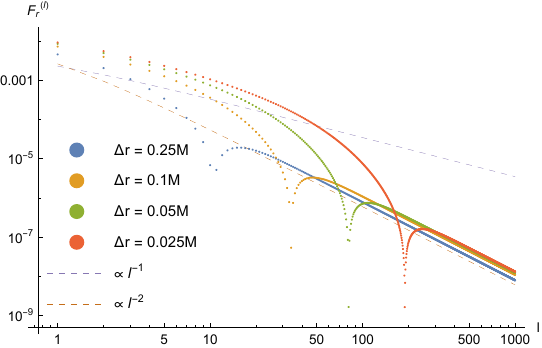}%
\label{boundary direct modes}}
\subfloat[Direct regularisation near the star boundary from the interior]{%
  \includegraphics[width = 0.5\textwidth]{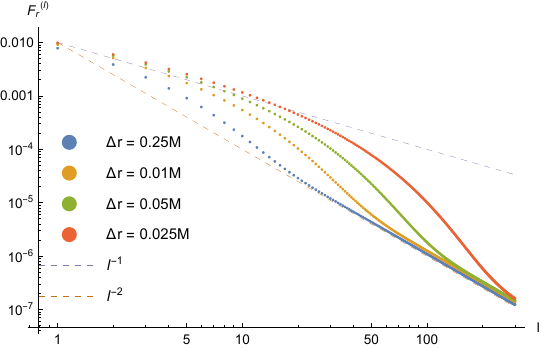}%
\label{int boundary direct modes}}
\vspace{5 px}
\subfloat[Difference regularisation near the star boundary]{%
  \includegraphics[width = 0.6\textwidth]{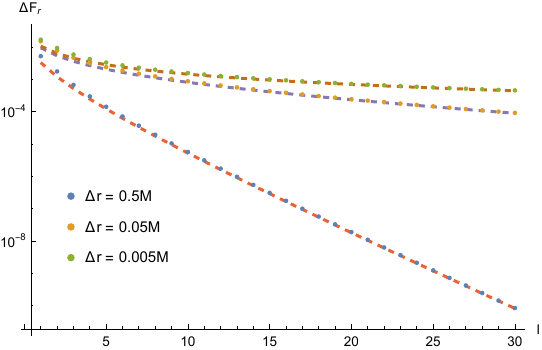}%
\label{boundary diff modes}}
\caption{The modes of the SF calculated via direct regularisation (upper) and via the difference method (lower) for a particle at $r_0 = R + \Delta r$, near the surface of a Schwarzschild star of radius $R=3M$. Plots (a) and (b) show that, as the particle gets closer to the surface of the star, the number of $\ell$ modes required to reach a regime where the bare modes show the expected $\left(\ell + 1/2\right)^{-2}$ fall off (in plots (a) and (b)) increases linearly with $1/|\Delta r|$. Plot (c) shows the difference-regularized $\ell$-modes, for a range of $\Delta r$. The dashed lines show the large-$\ell$ asymptotic approximation in Eq.~\eqref{surfexp}. The exponent $\Omega$ in that approximation approaches zero as $\Delta r \rightarrow 0$, and thus the total self-force diverges as $\Delta r \rightarrow 0$ in a logarithmic manner. 
}\label{verifying regularisation}
\end{figure}

It is interesting to examine how the self-force behaves as the charge (at $r=r_0$) approaches the surface of the star (at $r=R$), from either the interior or exterior direction. Naively, one can take the expressions for the $\ell$-modes in Eqs.~\eqref{starSF1}, \eqref{scalarregparam} and \eqref{intSF} and plot them for various values of $\Delta r = r_0 - R$. The behaviour of the $\ell$-modes is shown in Fig.~\ref{surfcompare} and Fig.~\ref{verifying regularisation}. 

In the case of the directly regularised modes, the plots show that there exists an intermediate regime $\ell \lesssim \lreg$ in which the modes fall as approximately $(\ell+1/2)^{-1}$, in addition to the asymptotic regime $\ell \gtrsim \lreg$ in which the regularized modes show the expected $(\ell+1/2)^{-2}$ fall off. 
Moreover, the value of $\lreg$ scales in inverse proportion to $\Delta r$, such that $\lreg \rightarrow \infty$ in the limit $\Delta r \rightarrow 0$. Figure \ref{boundary diff modes} (lower plot) also shows the situation for  \emph{difference} regularization. Again, there is an intermediate regime with an approximate $(\ell + 1/2)^{-1}$ scaling (for $\ell \lesssim \lreg$), and then the exponential decay takes over (for $\ell \gtrsim \ell_{\text{reg}}$). 

One can put forward a somewhat intuitive explanation for the scaling of $\ell_{\text{reg}}$ observed in Fig.~\ref{verifying regularisation}. The $S$ field, used for regularization, has a local definition and it is well-characterised by regularization parameters in either the Schwarzschild exterior or interior spacetime regions, but not across the star boundary. Near the surface of the star, we assume that the $S$ field is only properly defined by the regularization parameters in an open ball around the charge, in one of these two spacetime regions. That ball is of radius no larger than approximately $\Delta r$. The parameter $\ell$ may be considered as a measure of angular resolution on the sphere at fixed $r$, with $\Delta \theta \sim \pi / (\ell + 1/2)$ for large $\ell$. The closer the field point is to the the surface, the larger the $\ell$ value necessary to resolve the (small) open ball, and the consequently the larger the value of $\ell_{\text{reg}}$. From $\Delta r \sim R \Delta \theta$ we obtain the estimate $\lreg \sim \pi R / \Delta r$. This suggests that $\lreg$ diverges in the limit $\Delta r \rightarrow 0$, as observed in the numerics. Thus, for $\Delta r \rightarrow 0$, one has an infinite number of modes which fall as $(\ell+1/2)^{-1}$, and thus the self-force itself is (logarithmically) divergent in the approach to the surface. 

To understand this phenomenon more precisely, one can employ asymptotic expansions for the Legendre polynomials and Legendre functions used in the calculation of the SF \cite{NIST:DLMF} through difference regularization. Performing such an expansion for large $\ell$ (see Appendix \ref{sec:asymptotics} for details), we obtain the following expansion for the modes of the SF difference in the exterior: 
\begin{align}\label{surfexp}
F_r^{\ell} \sim F_r^{\ell \infty} &\equiv \f{q^2}{4M^2}\f{(Z + \delta  +1)^{-3/2}}{(Z+1) \sqrt{Z+\delta -1}}\f{e^{-\Omega \left(2\ell+1\right)}}{\ell+1/2}, \\
\Omega &\equiv \log\left({\frac{Z+\delta +\sqrt{(Z+ \delta )^2-1}}{Z+\sqrt{Z^2-1}}}\right).
\end{align}
In the above expression we take the charge to be at $z_0 = Z + \delta$ (for arbitrary $\delta = \Delta r / M$), where we continue to use the harmonic coordinate $z$ defined in \eqref{eq:zdef}.  

The asymptotic approximation in Eq.~\eqref{surfexp} exhibits the behavior described above, and seen in Fig.~\ref{verifying regularisation}. For $\ell + 1/2 \ll 2 / \Omega$, the modes fall off as $(\ell + 1/2)^{-1}$. Exponential decay takes over once $\ell + 1/2 \gtrsim 2 / \Omega$. Close to the surface ($\delta \ll 1$) the exponent is $\Omega \approx \delta / \sqrt{Z^2-1}$, and thus we recover the essentials of the scaling of $\lreg$ observed in Fig.~\ref{verifying regularisation}.

To derive a first approximation to the self-force near the surface of the star, we sum the modes $F_r^{\ell \infty}$ in the asymptotic approximation Eq.~\eqref{surfexp} to get
\begin{equation}\label{approxsum}
F_r^{\textbf{surf}} \equiv \sum_{\ell = 0}^\infty F_r^{\ell \infty} = 
\f{q^2}{4M^2}\f{(Z+\delta +1)^{-3/2}}{(Z+1) \sqrt{Z+\delta -1}}
\ln \left( \coth (\Omega / 2) \right) .
\end{equation}
This formula makes it clear that the self-force diverges logarithmically as $\delta \rightarrow 0$. 

\begin{figure}[ht]
    \centering
    \subfloat[Scalar SF across the full domain]{\includegraphics[width=0.49\linewidth]{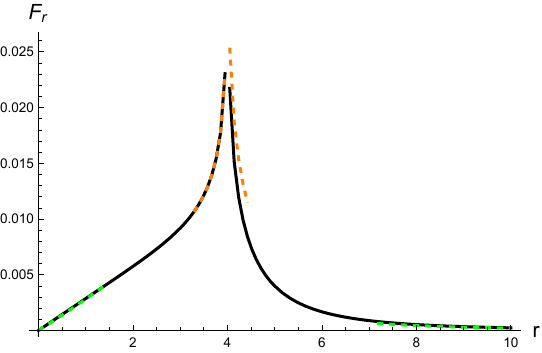}\label{sums}}
    \subfloat[Electromagnetic SF across the full domain]{\includegraphics[width=0.49\linewidth]{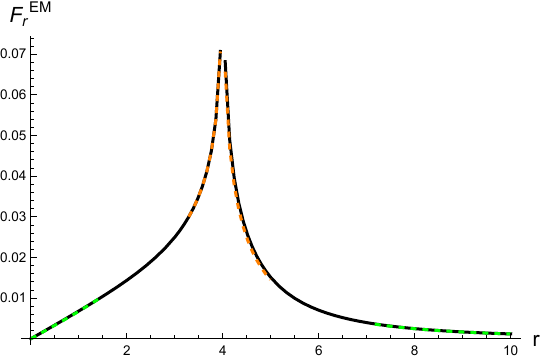}\label{EMsums}}
    \vspace{2 px}
    \subfloat[Verifying the logarithmic scaling of the SF with $\Delta r$]{\includegraphics[width=0.65\linewidth]{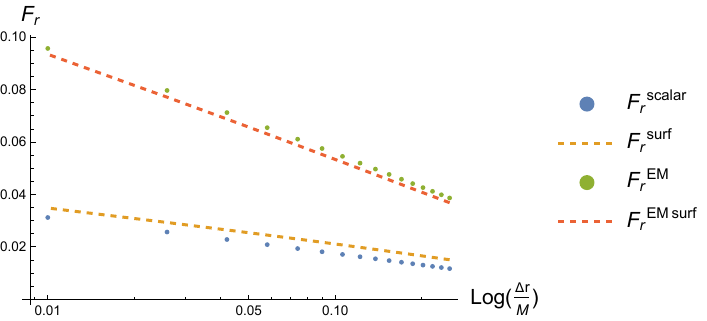}\label{llplots}}
    \caption{
  In Fig. \ref{sums} and Fig. \ref{EMsums}, we compare the scalar and the EM SF over the full radial domain. The solid lines represent a sum of the numerically-calculated $\ell$-modes, while the orange (green) dashed lines represent the approximation near the surface (near the centre and in the far field). A key difference between the two sums is the contribution of the monopole term in the scalar field sum which is absent in the EM case. Fig. \ref{llplots} compares the numerical sum with the approximations, again, but with a logarithmic scale for the $x$-axis. This confirms that the SF diverges in proportion to $\log{\left(\Delta r/M\right)}$ in the approach to the surface. All plots are for a star of radius $R = 4M$. 
    }
    \label{BoundarySF}
\end{figure}

Figure \ref{BoundarySF} shows the divergence of the self-force in the approach to the boundary of the star. For the scalar-field case, Fig.~\ref{sums} compares the numerical results from performing the mode sum with regularization, with the asymptotic result obtained in Eq.~\eqref{approxsum}. The comparison in Fig.~\ref{llplots} verifies that the divergence is of a logarithmic form for $\Delta r \rightarrow 0$. 

In the interior, one can also insert large-$\ell$ asymptotics for the mode functions into the self-force difference formulae in Eq.~\eqref{eq:Frdiff-interior}. Following the procedure above, we find a logarithmic divergence of a similar form to the above as the star's surface is approached from the interior. The expressions are rather long and are omitted here.

\subsubsection{Electromagnetic self force}
The electromagnetic SF is given by the SF difference mode-sum (Eq.~\eqref{EMsfdiff}) and an additive non-zero SF coming from the BH contribution (Eq.~\eqref{eq:FrBH}) which is independent of the star's radius. To understand the behaviour of the SF as one approaches the boundary of the star, it is sufficient to examine the former term. Looking at the large $\ell$ asymptotics of the Legendre functions once again, one has an asymptotic expansion for the EM SF difference. We write down the modes in the EM case in terms of the modes from the scalar case in Eq.~\eqref{surfexp}
\begin{equation}\label{EMsurfexp}
    \left(\Delta F_r^{EM}\right)^{\ell} \sim 3F_r^{\ell \infty}~.
\end{equation}
However, in this case one must consider the sum starting with the dipole term, since there is no monopole contribution in the EM SF difference. Thus, while there still exists a closed form expression for this infinite sum, it is not identical to that of the scalar field. The EM SF difference is given by $F_r^{\textbf{EM surf}}  = 3F_r^{\textbf{surf}} - 3F_r^{0 \infty}$, or more explicitly,
\begin{align}
    F_r^{\textbf{EM surf}} &= \f{q^2}{M^2}\frac{3 \left(\sqrt{(Z+\delta)^2-1)}+Z+\delta\right) \tanh ^{-1}\left(e^{-\Omega}\right)-3 \left(\sqrt{Z^2-1}+Z\right)}{2 (Z+1) \sqrt{\delta +Z-1} (\delta +Z+1)^{3/2} \left(\delta +\sqrt{(\delta +Z)^2-1}+Z\right)} \label{EMapproxsum} .
\end{align}
The approximation above is compared with numerical sums for the SF in Fig.~\ref{EMsums}. 

In the interior, we can also use the difference term to find an approximation for the force near the star's surface. In the harmonic coordinates $Z = R/M-1$ with the separation given in terms of $\delta = Z- z_0\geq 0$, we have the asymptotic expression: 
\begin{equation}\label{intEMasymptotics}
    \left(\Delta F_r^{\text{EM int}}\right)^\ell = -\frac{q^2\left(Z^2+(5-6 Z) \log ^2\left(\sqrt{Z^2-1}+Z\right)-1\right)}{8 M^2 \left(Z^2-1\right) \sqrt{1-\frac{2 (-\delta +Z+1)^2}{(Z+1)^3}} (-\delta +Z+1)^2 \log ^2\left(\sqrt{Z^2-1}+Z\right)}\f{e^{-\Omega(\ell+1/2)}}{\ell+1/2}
\end{equation}
where 
\begin{equation}
    \Omega = -2 \log \left(\frac{(Z+1)^{3/2} \left(\sqrt{1-\frac{2(Z-\delta+1)^2}{(Z+1)^3}}-1\right)}{\left(\sqrt{Z-1}-\sqrt{Z+1}\right) (Z-\delta+1)}\right)~.
\end{equation}
For small $\delta$, 
\begin{equation}
    \Omega \approx 2 \frac{ \left(1 - Z \left(\sqrt{\frac{Z+1}{Z-1}}-1\right)\right)}{(Z+1) \left(\sqrt{Z^2-1}-Z\right)} \delta . 
\end{equation}
Approaching the surface from the interior, we find that $\Omega \to 0$ as $\delta \to 0$, much like in the exterior. Hence the force diverges in proportion to the logarithm of $|r - R|$, whether one approaches the surface from the interior or the exterior of the star. 

\subsection{Across the radial domain\label{sec:full-domain}}

\begin{figure}
    \centering
    \subfloat[Scalar SF]{\includegraphics[width=0.8\linewidth]{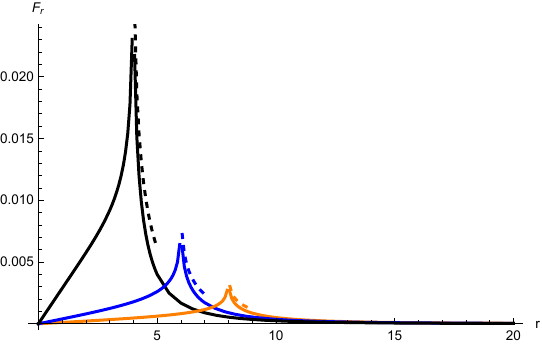}\label{FDscalar}} \\
    \subfloat[Electromagnetic SF]{\includegraphics[width=0.8\linewidth]{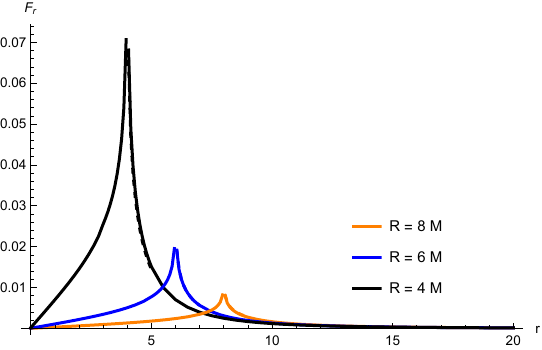}\label{FDEM}}
    \caption{Self force across the domain in $r$. We compare the force across different stellar radii and show the agreement between the approximate sums [dashed] with the numerical sums [solid] near the surface of the star. We see that the force tends to zero at the stellar centre and also as $r \to \infty$.  
    }
    \label{fig:fulldomainSF}
\end{figure}

Figure \ref{fig:fulldomainSF} shows the scalar SF across the entire domain, i.e.~for $r \in (0,R) \cup (R, \infty)$. The SF at the star boundary $r=R$ is undefined; the dashed lines show the approximation \eqref{approxsum} that captures the logarithmic divergence in the approach to the boundary. We note that the SF decreases with an increase in stellar radius.

\section{Discussion and conclusions\label{sec:conclusions}}

In the preceding sections we have computed the self-force acting on a pointlike particle endowed with a (electromagnetic or conformal-scalar) charge $q$, that is held in position in the vicinity of a Schwarzschild star (i.e.~a transparent sphere of constant density). We used two complementary methods for regularization, and we leveraged the fact that the Schwarzschild interior geometry is conformal to a three-sphere geometry, which in turn implies that dense Schwarzschild stars ($R \le 3M$) can produce perfect focussing, \emph{a la} Maxwell's fisheye lens. Via the means of a conformal transformation we obtained mode functions in closed form.  The key results comprise: series expansions in the large-$r$ regime in Eqs.~\eqref{scalarexp}--\eqref{eq:S1s0} and \eqref{EMexp}--\eqref{eq:S2s1} and the small-$r$ regime in Eqs.~\eqref{eq:Fcentre-s0} and \eqref{eq:Fcentre-s1}; approximations that describe the logarithmic divergence in the self-force at the star's boundary, Eq.~\eqref{approxsum} and \eqref{EMapproxsum}; and numerical data for the self-force across the full radial domain, as shown in Fig.~\ref{fig:fulldomainSF}. 

We have also extended the self-force difference method of Refs.~\cite{Drivas:2010wz, Isoyama_2012} to evaluate the self-force on a particle in the non-vacuum interior of the star, as well as in the vacuum exterior. This extension relies on the vanishing of certain regularized mode sums, given in Sec.~\ref{sec:identities}. This suggests that the self-force difference method of Drivas and Gralla \cite{Drivas:2010wz} may have validity to a wider range of circumstances than previously expected.

The \emph{electromagnetic} self-force near a Schwarzschild star was previously considered in Ref.~\cite{Shankar_2007} (see the ``insulating star'' case in that work). There, particular attention was given to a very compact star at exactly the Buchdahl radius, $R = R_{\text{Buch}} = 9M/4$, and it was reported that ``as we move the charge close to the star, $x \rightarrow 4/9$ [where $x = M/R$], $F_r$ becomes orders of magnitude greater than $F_r^{\text{BH}}$''. Here we have expanded on that observation, showing that the self-force diverges in logarithmic fashion as $r \rightarrow R$ for Schwarzschild stars of \emph{any} radius $R$. In other words, the divergence in the self-force is not some special feature of the Buchdahl limit. As in Ref.~\cite{Shankar_2007}, we used a conformal transformation (though of a different kind) to simplify the calculation of mode functions. The approach we took has the conceptual advantage of relating the interior Schwarzschild geometry to that of a three-sphere of constant curvature, but the practical disadvantage that our conformal transformation (unlike that in Ref.~\cite{Shankar_2007}) breaks down at the Buchdahl limit.

We have shown that the self-force for a Schwarzschild star is repulsive (from the star centre) across the entire domain in $r$ (i.e.~$F_r > 0$), in both the interior of the star and the Schwarzschild exterior. This suggests that, for a (loose) electrostatic analogy, one should consider the repelling force on a charge near an \emph{insulating} (rather than conducting) surface (i.e.~a surface boundary condition $\mathbf{n} \cdot \boldsymbol{\nabla} V(r)|_{r=R} = 0$).

It is interesting to compare our results for the self-force near a Schwarzschild star with another case that has been studied in the literature: the self-force on an electromagnetic charge held fixed inside a spherical shell of mass $M$ at $r=R$ (this was studied in Refs.~\cite{Unruh:1976fc,Burko_2000} and discussed in Ref.~\cite{Davidson:2018vqe}). At leading in order in $M/R$,
\begin{equation} \label{eq:F-mass-shell}
    \mathbf{F}_{\text{shell}} \approx -\frac{q^2 M}{2 R r^2} \left( \frac{r / R}{1 - r^2 / R^2} + \frac{1}{2} \ln\left(\frac{1 - r / R}{1 + r / R} \right) \right) \hat{\mathbf{r}}
\end{equation}
This force exhibits a $1/\Delta r$ divergence in the approach to the shell's surface, and the charge is repelled from the shell. This is in contrast to the (weaker) logarithmic divergence found for the Schwarzschild star. Near the centre of the mass-shell there is a restoring force of $\mathbf{F}_{\text{shell}} \approx -\mathbf{r} q^2 M / (3 R^4)$ that generates Simple Harmonic Motion. By comparison, in the Schwarzschild star the force is also proportional to $r$ in magnitude (see Fig.~\ref{fig:fulldomainSF}, and Eqs.~\eqref{eq:Fcentre-s0} and \eqref{eq:Fcentre-s1}), but the force is in the opposite direction (i.e.~repelling the particle from the centre) and so SHM does not arise.

Another scenario where a comparison can be drawn is to the self-force on a particle in a spacetime constructed by glueing together two Minkowski spacetimes along a spherical seam \cite{Davidson:2018vqe}. The Riemann tensor is zero everywhere except at the seam, where is proportional to a delta-distribution. In this scenario, a $1/\Delta r$ divergence in the self-force arises in the approach to the seam.

It seems that divergences in the self-force are rather typical wherever there is a boundary, or a non-smooth feature of the geometry or refractive index. One may arrange such examples into a \emph{hierarchy of divergences}, based on the scaling of the force with $\Delta r$ in the approach to the boundary. In familiar electrostatic examples, the force on a charge typically diverges as $1/(\Delta r)^2$. For example, near a conducting (or insulating) plate or sphere; or at a boundary where there is a jump in the refractive index (e.g.~outside or inside a dielectric sphere). In the curved-spacetime example of a mass shell (Eq.~\eqref{eq:F-mass-shell}) \cite{Unruh:1976fc,Burko_2000}, or of glued-Minkowski \cite{Davidson:2018vqe}, the divergence instead scales as  $1/(\Delta r)$, i.e., it is one power weaker. In the Schwarzschild-star case, we found a logarithmic divergence only, which can be regarded as one power weaker again. 

It is natural to seek to relate the form of the divergence to the smoothness of the geometry. Let us examine this idea in a little more detail. For the Schwarzschild star geometry, the metric functions $A(r)$ and $B(r)$ are continuous at $r=R$. More precisely, $A(r)$ is $C^1$ (continuous and once-differentiable) and $B(r)$ is $C^0$.
One point of comparison is the force on a pointlike charge outside a dielectric sphere in electrostatics. Here,  there is a discontinuity in the refractive index at $r=R$ (i.e.~it is not $C^0$) which leads, via the method of images, to a $(\Delta r)^{-2}$ divergence in the force. For the Schwarzschild star, the effective refractive index is $C^1$ at the boundary, that is, \emph{two} orders smoother than for a dielectric sphere. It seems plausible that the divergence in the self-force, with $\log |\Delta r|$, is two orders weaker than in the dielectric case simply because the (effective) refractive index is two orders more differentiable.

Due to the divergence at the star surface, the self-force on a pointlike particle near a Schwarzschild star has no upper bound; it can be arbitrarily large. The divergence is likely to be an artifact of the point-particle assumption. For an extended body, one would expect the force to remain bounded, because a logarithmic divergence is weak enough to be integrable. 

In the static scenario considered here, the self-force is entirely \emph{conservative} (i.e.~symmetric under time reversal). In Ref.~\cite{Drivas:2010wz} it was posited that the dissipative (radiation-reaction) part of self-force (i.e.~the part that drives the loss of energy/angular momentum to radiation in dynamical situations) should have a \emph{local} character, whereas the conservative part of the self-force is sensitive to boundary conditions and the global structure of the spacetime. Consequently, the latter can become large while the former remains small (or zero). Our results are consistent with this hypothesis, since we have shown that the conservative (point particle) self-force is unbounded in the approach to a boundary where the metric itself is continuous and the Ricci tensor is bounded.

\begin{acknowledgments}
S.~D.~acknowledges financial support from the Science and Technology Facilities Council (STFC) under Grant No.~ST/X000621/1 and Grant No.~ST/W006294/1. A.~N.~S.~is supported by an EPSRC Ph.D studentship. 
\end{acknowledgments}

\appendix

\section{Large $\ell$ expansions}\label{sec:asymptotics}
We want to perform large $\ell$ asymptotics for the following Legendre functions: $P_{\ell}(x), Q_{\ell}(x)$ when $x>1$ and $P_{\pm1/2}^{-\ell-1/2}(x)$ when $-1\leq x \leq 1$. From Sec.~14.15 in \cite{NIST:DLMF} and \cite{olver1997asymptotics} we have: 
\begin{align}
    P_\ell(\cosh{\xi}) &\sim \left(\f{\xi}{\sinh{\xi}}\right)^{1/2}\left\{I_0\left(L\xi\right)\sum_{s=0}^{p}\f{A_s^{0}\left(\xi^2\right)}{L^{2s}} + \f{\xi}{L}I_{-1}\left(L\xi\right)\sum_{s=0}^{p-1}\f{B_s^{0}\left(\xi^2\right)}{L^{2s}}\right\} \\
    Q_\ell(\cosh{\xi}) &\sim \left(\f{\xi}{\sinh{\xi}}\right)^{1/2}\left\{K_0\left(L\xi\right)\sum_{s=0}^{p}\f{A_s^{0}\left(\xi^2\right)}{L^{2s}} - \f{\xi}{L}K_{1}\left(L\xi\right)\sum_{s=0}^{p-1}\f{B_s^{0}\left(\xi^2\right)}{L^{2s}}\right\}|,
\end{align}
where $L = \ell+1/2$ and $I_0, I_{-1}, K_0, K_{-1}$ are modified Bessel functions of the first and second kind with degree $0,-1$ respectively. For our purposes the summations in both expansions can be truncated at order $s=0$ with the leading coefficients given by $A_0^{0} = 1$ and 
\begin{equation*}
    B_0^{0}\left(\xi^2\right) = \f{1}{8\xi}\left(\coth{\xi}-\f{1}{\xi}\right).
\end{equation*}
The Bessel functions have large argument expansions given by \cite{NIST:DLMF}, 
\begin{equation}
    I_{\nu}(x) \sim \f{e^{x}}{\left(2\pi x\right)^{1/2}}\sum_{k=0}^{\infty}\left(-1\right)^k\f{a_k\left(\nu\right)}{x^k}, \quad K_{\nu}(x) \sim \left(\f{\pi}{2x}\right)^{1/2}e^{-x}\sum_{k=0}^{\infty}\f{a_k\left(\nu\right)}{x^k}, 
\end{equation}
with $a_{0}(\nu) = 1$ and,
\begin{equation*}
    a_{k}(\nu) = \f{\left(\f{1}{2}-\nu\right)_k\left(\f{1}{2}+\nu\right)_k}{(-2)^kk!} \hspace{20 px} \quad k \geq 1.
\end{equation*}
Finally, the Legendre functions, which go into the field solutions in the interior have the expansion for $-1 \leq x \leq 1$, 
\begin{equation}
    P_{\nu}^{-\mu}(x) = \left(\f{1- x}{1 + x}\right)^{\mu/2}\left\{\sum_{j=0}^{J-1}\f{\left(\nu+1\right)_j\left(-\nu\right)_j}{j! \Gamma\left(j+1+\mu\right)}\left(\f{1-x}{2}\right)^j + \mc{O}\left(\f{1}{\Gamma\left(J+1+\mu\right)}\right)\right\}.
\end{equation}
where the non-negative integer $J$ denotes the order of expansion. In all the above expansions, $\left(y\right)_k$ are the Pochhammer symbols. We use these expansions to see how the modes of the SF difference behave at large $\ell$ and the most interesting aspect of this comes from the structure factor $S_{\ell}(Z)$. At leading order in the large $\ell$ expansions the $\ell$ behaviour of the modes entirely depends on the $\ell$ behaviour of $S_{\ell}(Z)$. We recall from (\ref{starSF2}) the SF difference now rewritten for convenience, 
\begin{equation}
    \Delta F_{r}^{\ell} =  -  \left(\f{q}{M}\right)^2\sqrt{\f{z_0-1}{z_0+1}}\sum_{\ell}\left(2\ell+1\right)Q_{\ell}(z_0) Q_{\ell}'(z_0)~\left\{\f{\left(Z+1\right)\f{P_{\ell}'(Z)}{P_{\ell}(Z)} - \eta_{\ell}(Z)}{\left(Z+1\right)\f{Q_{\ell}'(Z)}{Q_{\ell}(Z)} - \eta_{\ell}(Z)}\right\}\f{P_{\ell}(Z)}{Q_{\ell}(Z)} 
\end{equation}
with the explicit structure factor in the braces. For the two terms in the numerator of the structure factor, we arrive at the following asymptotic expansions from above: 
\begin{align}
    \left(Z+1\right)\f{P_{\ell}'(Z)}{P_{\ell}(Z)} &\sim \frac{L (Z+1)}{\sqrt{Z^2-1}}+\frac{Z}{2-2 Z}+\frac{1}{L (8-8 Z) \sqrt{Z^2-1}}+\frac{Z}{64 L^2 (Z-1)^2 (Z+1)}+\mc{O}\left(\frac{1}{L^3}\right) \\
    \eta_{\ell} &\sim L \sqrt{\frac{Z+1}{Z-1}}+\frac{Z}{2-2 Z}+\frac{4 Z-5}{8 L (Z-1)^{3/2} \sqrt{Z+1}} \nonumber \\
    &+\frac{(4 Z-5) \left(Z \left(\sqrt{Z^2-1}-9 Z\right)-2 \sqrt{Z^2-1}+9\right)}{64 L^2 (Z-1)^{5/2} (Z+1)^{3/2}}+\mc{O}\left(\frac{1}{L^3}\right).
\end{align}

We can now see that the terms in the structure factor at order $L^1$ and $L^0$ cancel, causing the SF difference to vanish at these orders. Thus, it is only at order $L^{-1}$ that we expect a non-zero contribution to the structure factor. 

    \section{A vanishing mode sum}\label{sec:proof}
\textbf{Proposition: } The following infinite sum is zero:
\begin{equation}
 \sum_{\ell=0}^{\infty} \left\{ (2 \ell + 1) P_{\ell}(z) Q_{\ell}(z) - \gamma_0(z) \right\} = 0 ,
\end{equation}
where $P_{\ell}(z)$ and $Q_{\ell}(z)$ are Legendre functions ($z > 1$), and the regularization function is $\gamma_0(z) \equiv \frac{1}{\sqrt{z^2 - 1}}$,

\textbf{Proof: }
The starting point is Christoffel's second formula (14.18.7 in DLMF \cite{NIST:DLMF}), which is an identity for a partial sum from $0$ to $n \in \mathbb{N}$, viz.,
\begin{equation}
    \left(z-y\right) \sum_{l=0}^n \left(2\ell+1\right)P_{\ell}(z)Q_{\ell}(y) = \left(n+1\right)\left(P_{n+1}(z)Q_n(y)-P_n(z)Q_{n+1}(y)\right) - 1 \equiv F_n(z,y) ~,
\end{equation}
(This can be established by using the recurrence relations for Legendre functions and proof by induction). Since the partial sum on the left-hand side is finite, it is clear that the limit of $F_n(z,y) / (z-y)$ as $y \rightarrow z$ must be well-defined and finite. We evaluate this limit with l'H\^opital's rule to establish that 
\begin{equation}
\sum_{l=0}^n \left(2\ell+1\right)P_{\ell}(z)Q_{\ell}(z) = (n+1) \left( P_{n+1}^{\prime}(z) Q_{n}(z) - P_{n}^{\prime}(z) Q_{n+1}(z) \right).
\label{eq:PQpartialsum}
\end{equation}
Using the asymptotics of the previous section, one can then establish that, in the large-$n$ regime,
\begin{equation} \label{eq:asymptotic-result}
P_{n+1}^{\prime}(z) Q_{n}(z) - P_{n}^{\prime}(z) Q_{n+1}(z) = \gamma_0(z) + \mathcal{O}\left(\frac{1}{n^2} \right),
\end{equation}
where $\gamma_0(z)$ was defined above. Now subtracting $(n+1) \gamma_0(z)$ from both sides of \eqref{eq:PQpartialsum} gives 
\begin{equation}
\sum_{l=0}^n \left\{ \left(2\ell+1\right)P_{\ell}(z)Q_{\ell}(z) - \gamma_0(z) \right\} = (n+1)\left(P_{n+1}^{\prime}(z) Q_{n}(z) - P_{n}^{\prime}(z) Q_{n+1}(z) - \gamma_0(x) \right).
\end{equation}
Finally, taking the limit $n \rightarrow \infty$ and using the asymptotic result \eqref{eq:asymptotic-result} proves the proposition.

\section{The partial sum Eq.~\eqref{eq:interior-sum}} \label{sec:proof2}
In this section we show that Eq.~(\ref{eq:interior-sum}) follows as a consequence of the recurrence relation (DLMF 14.10.1 \cite{NIST:DLMF})
\begin{equation}
\label{eq:recurrence}
P_{\nu}^{\mu+2}(x) + 2 (\mu + 1) \hat{y}  P_{\nu}^{\mu+1}(x) + (\nu - \mu)(\nu + \mu + 1) P_\nu^\mu(x) = 0,  \quad \quad \hat{y} \equiv \frac{x}{\sqrt{1-x^2}} .
\end{equation}
A similar proof can be constructed for Eq.~\eqref{eq:PQpartialsum} and Eq.~\eqref{eq:interior-sum}. 

First, let us assume that the partial sum in Eq.~(\eqref{eq:interior-sum}) will take the form
\begin{equation}
\sum_{\ell = 0}^{n} \mathcal{S}_{(1)\ell}^{\text{int.}}(x) = \mathfrak{S}_n(x) + F_0(x) , \label{eq:ansatz}
\end{equation}
where $\mathcal{S}_{(1)\ell}^{\text{int.}}(x)$ is defined in Eq.~\eqref{eq:S1int}, $F_0(x) \equiv \mathcal{S}_{(1)\ell}^{\text{int.}} - \mathfrak{S}_0$ and we posit that $\mathfrak{S}_n(x)$ takes a form inspired by \eqref{eq:PQpartialsum}:
\begin{equation}
\mathfrak{S}_n(x) = \beta_n \left( P_{1/2}^{-N}(x) \partial_{\hat{y}} P_{1/2}^{N+1}(x) - \alpha_n P_{1/2}^{-N-1}(x) \partial_{\hat{y}} P_{1/2}^{N}(x) \right) , \label{eq:frakS}
\end{equation}
where $N=n+1/2$, and $\alpha_n$ and $\beta_n$ are to be determined below. 
From Eq.~\eqref{eq:recurrence} it follows that
\begin{align}
P_{1/2}^{N+2}(x) + (2 n + 3) \hat{y} P_{1/2}^{N+1}(x) - n(n+2) P_{1/2}^{N}(x) = 0, \label{eq:LegPrecurrence1} \\
P_{1/2}^{-N}(x) - (2 n + 3) \hat{y} P_{1/2}^{-N-1}(x) - (n+1)(n+3) P_{1/2}^{-N-2}(x) = 0. \label{eq:LegPrecurrence2}
\end{align}
 Using \eqref{eq:LegPrecurrence2} and the derivative of \eqref{eq:LegPrecurrence1} in \eqref{eq:frakS}, one can show that
\begin{align}
\mathfrak{S}_{n+1}(x) &= \beta_{n+1} P_{1/2}^{-N-1} \left[-(2n+3) \hat{y} \partial_{\hat{y}} P_{1/2}^{N+1} + n(n+2) \partial_{\hat{y}} P_{1/2}^{N} - (2 n + 3) P_{1/2}^{N+1} \right] \nn \\ 
& - \frac{\beta_{n+1} \alpha_{n+1}}{(n+1) (n+3)} \partial_{\hat{y}} P_{1/2}^{N+1} \left[ 
-(2 n + 3) \hat{y} P_{1/2}^{-N-1} + P_{1/2}^{-N}
\right] .
\end{align}
Now we choose $\alpha_{n+1} = (n+1)(n+3)$ so that a pair of terms cancel. After this cancellation,
\begin{align}
\mathfrak{S}_{n+1}(x) &=
\beta_{n+1} \left[ \alpha_{n+1} P_{1/2}^{-N-1} \partial_{\hat{y}} P_{1/2}^{N} - P_{1/2}^{-N} \partial_{\hat{y}} P_{1/2}^{N+1} \right] - \beta_{n+1} (2 n + 3) P_{1/2}^{-N-1} P_{1/2}^{N+1} . 
\label{eq:frakSplus1}
\end{align}
After comparing \eqref{eq:frakSplus1} with \eqref{eq:frakS}, we may choose $\beta_{n+1} = (-1)^{n}$ so that
\begin{align}
\mathfrak{S}_{n+1}(x)
 &= \mathfrak{S}_{n}(x) + \mathcal{S}_{(1)n+1}^{\text{int.}}(x) .
\end{align}
Hence the result for the partial sum \eqref{eq:ansatz} is established, and consideration of the base case ($n=0$) yields $F_0(x) = -x / \pi$. The result in Eq.~\eqref{eq:interior-sum} then follows after rewriting the derivative using the chain rule:  $\partial_{\hat{y}} = (1 - x^2)^{3/2} \partial_x$.

\clearpage 
\bibliographystyle{IEEEtran}
\bibliography{paperref.bib}

\end{document}